\documentclass[manuscript]{aastex}
\usepackage{natbib}
\shorttitle{Horseshoe drag in isothermal disks}
\shortauthors{J. Casoli \& F. S. Masset}

\begin{document}

\title{On the horseshoe drag of a low-mass planet. I - Migration in
  isothermal disks}
\author{J. Casoli\altaffilmark{1}}
\affil{Laboratoire AIM,
 CEA/DSM - CNRS - Universit\'e Paris Diderot, 
 Irfu/Service d'Astrophysique, B\^at. 709, 
 CEA/Saclay, 91191 Gif-sur-Yvette, France}
\email{jules.casoli@cea.fr}
\altaffiltext{1}{Send offprint requests to jules.casoli@cea.fr}
\and 
\author{F.\ S.\ Masset\altaffilmark{2}} 
\affil{Laboratoire AIM,
 CEA/DSM - CNRS - Universit\'e Paris Diderot, 
 Irfu/Service d'Astrophysique, B\^at. 709, 
 CEA/Saclay, 91191 Gif-sur-Yvette, France}
\email{frederic.masset@cea.fr}
\altaffiltext{2}{Also at ICF-UNAM, Av. Universidad s/n,
Cuernavaca, Morelos, C.P. 62210, M\'exico}
\begin{abstract}
  We investigate the unsaturated horseshoe drag exerted on a low-mass
  planet by an isothermal gaseous disk. In the globally isothermal
  case, we use a formalism, based on the use of a Bernoulli invariant,
  that takes into account pressure effects, and that extends the
  torque estimate to a region wider than the horseshoe region. We find
  a result that is strictly identical to the standard horseshoe
  drag. This shows that the horseshoe drag accounts for the torque of
  the whole corotation region, and not only of the horseshoe region,
  thereby deserving to be called corotation torque.

  We find that evanescent waves launched downstream of the horseshoe
  U-turns by the perturbations of vortensity exert a feed-back on the
  upstream region, that render the horseshoe region asymmetric. This
  asymmetry scales with the vortensity gradient and with the disk's
  aspect ratio.  It does not depend on the planetary mass, and it does
  not have any impact on the horseshoe drag. Since the horseshoe drag
  has a steep dependence on the width of the horseshoe region, we
  provide an adequate definition of the width that needs to be used in
  horseshoe drag estimates.

  We then consider the case of locally isothermal disks, in which the
  temperature is constant in time but depends on the distance to the
  star. The horseshoe drag appears to be different from the case of a
  globally isothermal disk.  The difference, which is due to the
  driving of vortensity in the vicinity of the planet, is intimately
  linked to the topology of the flow. We provide a descriptive
  interpretation of these effects, as well as a crude estimate of the
  dependency of the excess on the temperature gradient.
\end{abstract}


\section{Introduction}

Planetary migration has emerged in the last decades as one of the key
processes that shape forming planetary systems. Since the early work
of \citet{gt79,gt80}, who gave the expressions of the tidal torque between
a disk and a perturber at an isolated resonance, the tidal torque between
a protoplanet and the gaseous disk has been investigated in great detail.
In particular, \citet{ww86,WW88} worked out the sum of the torques at all
Lindblad resonances (the so-called differential Lindblad torque) and found that forming
planets should undergo a fast decay towards their central object.
Apart from the differential Lindblad torque, the total tidal torque between a planet and a disk features
another component,  the corotation torque, which arises from
material located in the vicinity of the orbit. Most of the early works
on planetary migration focused on the Lindblad torque, as it was
considered that the latter dominates the corotation torque, and
therefore dictates the direction of the migration and the order of
magnitude of the drift rate.

Recent works have nevertheless shown the importance of the corotation
region on the total torque exerted by the disk on the planet.
Traditionally, it was estimated, for low mass planets, by the use of a
linear theory.  This amounts to evaluating the amplitude of the
resonant waves at different azimuthal wave-numbers excited by the
planet in the corotation region, and their feed back on the
planet. Such an approach was adopted by numerous authors, such as
\citet{1989ApJ...336..526W}, \citet{1993Icar..102..150K}, and, more
recently, by \citet{tanaka2002}.  Recently, however,
\citet{2009arXiv0901.2265P} have stressed the fact that the corotation
torque acting on a low mass planet is not correctly accounted for by a
linear estimate, except at early stages after a planet is ``turned
on'' in a disk.  After a relatively short time scale, the linear
estimates break down, and these authors show that the total torque
seems to be better described by the sum of the differential Lindblad
torque and of the so-called horseshoe drag.  This component of the
torque arises from particles that are close to the corotation and
undergo a U-turn in front of or behind the planet. By doing this, they
exchange angular momentum with the planet, exerting a torque on it. As
the region in which these particles librate is called the horseshoe
region, the resulting torque was called the horseshoe drag. It is
evaluated by budgeting the jumps of angular momentum of the particles
undergoing a U-turn. This was first studied by \citet{wlpi91}, and
more recently by \citet{masset01,masset02,mp03,2009arXiv0901.2265P}.
This raises the question of the distribution of the torque density in
the coorbital region. While a linear analysis shows that the
corotation torque arises from an annulus of radial width $\sim H$
around corotation, the horseshoe drag approach considers only the
horseshoe region, which can be arbitrarily thin, and whose width
formally tends to zero as the planetary mass does. One can therefore
wonder if there is a torque exerted by the material close to the
orbit, but outside of the horseshoe region, and more generally how to
connect the linear analysis, which considers the launch of evanescent
waves in the coorbital region, and the horseshoe drag, which does not
contemplate them. Also, since the coorbital region and horseshoe
region become of increasing importance in studies of planetary
migration, it is useful to derive a number of properties of this
region.

With this in mind, we present in this paper~I an approach, valid in a
globally isothermal disk, in which we use a Bernoulli invariant to
handle the streamlines of the flow both in the horseshoe region and
outside of it, and in which we take into account the evanescent
pressure waves that arise on the downstream sides of the horseshoe
region, and which can act back on the upstream regions. We show that
taking these waves into account does not modify the horseshoe drag
estimate, but that they induce a rear-front asymmetry of the horseshoe
region, that scales with the vortensity gradient. We also work out the
surface density response in the corotation region, and we indicate a
proper way of estimating the horseshoe region width. This is critical
for horseshoe drag estimates based on streamline analysis, since this
drag scales with the fourth power of the horseshoe width.  Note that
in paper~II, we present a derivation of the horseshoe drag in an
adiabatic disk. We use a different invariant to derive it, but parts
of the derivation are very similar to the derivation undertaken in the
present paper.

We also propose a basic model of the mechanisms taking place in the
locally isothermal case, when the disk's temperature depends on the
radius.  Namely, there is creation of vortensity which takes place
near the planet, and which results in a corotation torque excess that
differs from the globally isothermal case, and which depends on the
temperature gradient.  Much like in the adiabatic case considered in
paper~II, this torque excess comes from an edge term of the horseshoe
drag. Nevertheless, owing to the lack of an invariant along the
streamlines in the locally isothermal case, we only provide a rough
estimate of the excess, and discuss the differences between the
predictions of this estimate and the results of numerical simulations.

\section{Basics}

\subsection{Problem definition and notation}
\label{sec:notations}

The system we study is composed of a central star, of mass $M_\ast$,
and, orbiting around it, a gaseous disk, with an embedded low-mass
planet, of mass $M_p$, which has a fixed circular orbit of radius
$a$ and an angular frequency $\Omega_p$. We assume the disk to be
thin, and we use a 2D cylindrical geometry, averaging or integrating
all the quantities over $z$.

We place ourselves in the frame corotating with the planet, in which
we assume the flow to have reached a steady state.  Nevertheless, we
explicitly discard in this work possible saturation effects.  This
amounts to assuming that the material which arrives at the upstream
side of the horseshoe U-turns does so with the vortensity of the
unperturbed disk.  In practice, this means that we consider the flow
after a time longer than the time required to complete the horseshoe
U-turns, but shorter than half the horseshoe libration time. 
In a real situation, the horseshoe flow achieves many librations over the disk's lifetime,
and the horseshoe drag saturates to a level that depends on the dissipative
processes at work in the disk \citep[e.g.\ ][]{masset01}. Hereafter we restrict
ourselves to the unsaturated horseshoe drag. The saturation properties,
both in the isothermal and adiabatic cases, will be considered in a forthcoming work.

The angular frequency of a fluid element in the disk at a distance $r$
of the star is $\Omega(r)$. The corotation radius, $r_c$, is the
radius where the gas is rotating at the same speed as the planet:
$\Omega(r_c)=\Omega_p$.  The disk is made of an ideal gas, and is
globally isothermal: its temperature is constant in time and
space. Initially, the surface density is a power law of the radius:
\begin{equation}
\Sigma = \Sigma_c \left(
  r/r_c \right)^{-\sigma},
\end{equation}
where $\Sigma_c$ is the surface density at corotation.

We denote the height of the disk by $H$, and we use the aspect ratio
$h=H/a$. As the disk is globally isothermal, the sound speed $c_s$
is constant across it, and the aspect ratio is only a function of $r$:
$h=c_s/[a\Omega(r)]$.

We make use of the vortensity $w=\omega/\Sigma$, where $\omega$ is the
vertical component of the flow's vorticity: $\omega=\vec{\nabla}
\times \vec{v}|_z$. With our notation, we have, initially: $ w = w_c
\left( r/r_c \right)^{\mathcal{-V}} $, with $ {\mathcal{V}}=3/2 -
\sigma$.

The particles feel the gravitational potential
$\Phi=\Phi_\ast+\Phi_p+\Phi_i$, where $\Phi_\ast=-GM_\ast/r$ is the
star's potential, $\Phi_p$ the potential of the planet, softened to
avoid computational problems:
\begin{equation}
\Phi_p=-GM_p/\left(r^2-2ra\cos\phi+a^2+\epsilon^2\right)^{1/2},
\end{equation}
(where $\epsilon$ is the softening length, typically $1/3$ of $H$),
and where $\Phi_i$ is the potential's indirect term:
\begin{equation}
  \label{eq:1}
  \Phi_i = \frac{GM_p}{a^2}r\cos\phi=qa\Omega_p^2r\cos\phi,
\end{equation}
where $q=M_p/M_*$.

The enthalpy is denoted $\eta$. For a globally isothermal disk of
temperature $T_c$, we have $\eta=T_c \ln(\Sigma/\Sigma_c)$.

We make use of the two Oort's constants, estimated at the planet's 
location, $A_p$ and $B_p$. We often use their values for a 
Keplerian disk, which reads : $A_p=-3\Omega_p/4$ and 
$B_p=\Omega_p/4$. The $p$ subscript is mandatory to distinguish
$B_p$ from the Bernoulli invariant, a quantity we define in 
section~\ref{sec:BernDef}.

\begin{figure}
	\centering
		\plotone{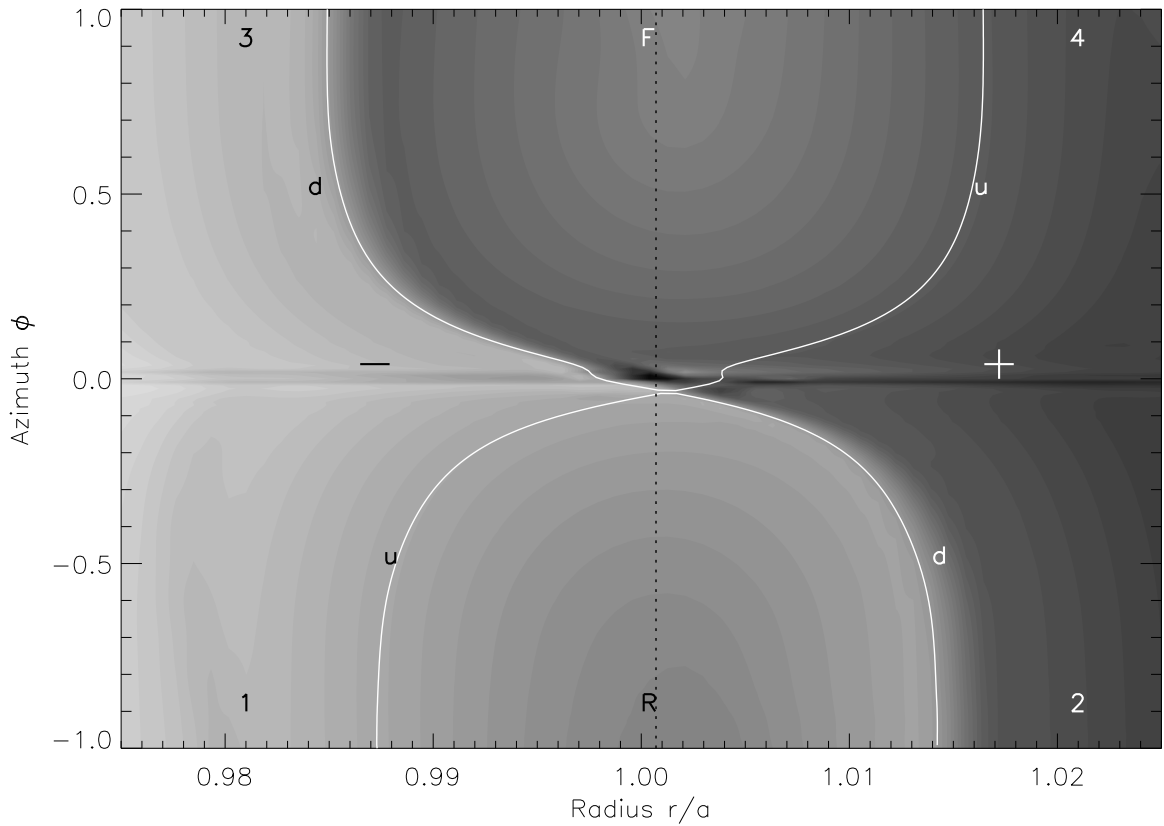}
                \caption{An overview of the horseshoe region. This is
                  extracted from the standard simulation (see
                  \ref{sec:Numeric}), at $T= 25$ orbits. The planet is
                  at $\phi=0$ and $r=1$. The corotation radius is
                  represented by the dotted line. The grey scale
                  represent the total vortensity, and the two white
                  lines are the separatrices. The asymmetry appears
                  clearly, the front part being wider than the rear
                  one. We also show here the integration domains 1 to
                  4 used in the text.}
	\label{fig:hsregion}
\end{figure}

We label the different quadrants of the horseshoe region as
represented in figure \ref{fig:hsregion}. We denote with a subscript
$F$ ($R$) what is in front (at the rear) of the planet,
rotation-wise. In a similar manner, we use a subscript $+$ ($-$) for
the part of the disk that has $r>r_c$ ($r<r_c$). Lastly, the upstream
(downstream) part of a horseshoe leg is denoted with a subscript $u$
($d$).

\subsection{Basic equations}
The governing equations are the one of a non magnetized, non
self-gravitating, ideal isothermal gas, in the corotating frame, so
that the radial velocity $v_r$ is unchanged with respect to the
inertial frame, while the angular velocity is $v_\phi = r\Omega -
r\Omega_p$.  With this notation the continuity equation reads:
\begin{equation}
  \label{eq:masscons}
  \partial_t\Sigma+\frac 1r\partial_r(\Sigma rv_r)+\frac
  1r\partial_\phi(\Sigma v_\phi)=0,
\end{equation}
while the Euler equations read, respectively in the radial and
azimuthal directions:
\begin{eqnarray}
\label{eq:eulerrad}
\partial_t v_r + v_r\partial_r v_r + \frac{v_\phi}{r} \partial_\phi v_r 
- r\Omega_p^2-2\Omega_p v_\phi - \frac{v_\phi^2}{r} &=&
-\frac{\partial_r P}{\Sigma} -\partial_r \Phi,\\
  \label{eq:euleraz}
  \partial_t j + v_r\partial_r j + \frac{v_\phi}{r} \partial_\phi j &=& 
  -\frac{\partial_\phi P}{\Sigma} -\partial_\phi\Phi,
\end{eqnarray}
where $j=r^2\Omega$ is the specific angular momentum.

\section{Horseshoe drag}
\subsection{A first torque expression}
To evaluate the torque exerted on the planet by the disk, we simply
budget the angular momentum on a well-chosen domain $D$, located
between two azimuths $\phi_D$ and $-\phi_D$, chosen far enough from
the planet to assume that we can neglect $\Phi_p$ there. The radial
extent of $D$ is bound by two streamlines, which respectively start at
$(r_c+x_D;\phi_D)$ and $(r_c-x_D;\phi_D)$. We assume that $x_D$ is
much larger than $x_s$, the typical half-width of the horseshoe
region.  The torque exerted by the material inside this region onto
the planet is then:
\begin{equation}
\label{eq:2}
\Gamma_{\rm disk \rightarrow planet} = 
\int\!\int_D \Sigma \partial_\phi \Phi \ rdr d\phi, 
\end{equation}
which can be recast, using (\ref{eq:euleraz}), as:
\begin{equation}
\label{eq:3}
\Gamma_{\rm disk \rightarrow planet} = \int\!\int_D \Sigma 
\left(- \left(\partial_t j + v_r\partial_r j + 
    \frac{v_\phi}{r} \partial_\phi j\right) - 
  \frac{\partial_\phi P}{\Sigma} \right) \ rdr d\phi, 
\end{equation}
which we note $\Gamma$ from now on. In steady state, we have, using
(\ref{eq:masscons}):
\begin{eqnarray}
\label{eq:4}
\Gamma &=& \int\!\int_D \frac{1}{r} \left(-\partial_\phi(v_\phi\Sigma j)
  -\partial_r(r\Sigma v_r j) - \partial_\phi
  rP \right) \ rdrd\phi \\
  \label{eq:gammaclosed}
  &=&-\oint_{\partial D}  \Sigma j \vec{v} \cdot \vec{dn} -
  \oint_{\partial D} rP\vec{e_\phi} \cdot \vec{dn},
\end{eqnarray}
where $\vec{dn}$ is a vector perpendicular to the edge $\partial D$
of the domain $D$, oriented outwards, and with a length equal to the
length of the elementary interval of integration on the edge.
To proceed in the treatment of this expression, we must rely on the
shape of the streamlines. Namely, for $r-r_c=\pm
x_D$, the streamlines are nearly circular and there is therefore no
advected flux of angular momentum into the domain $D$ through the
circular boundaries. We also use the fact that $\vec{e_\phi}\cdot
\vec{dn}=0$ on these boundaries, so that we can ultimately write:
\begin{equation}
  \label{eq:5}
  \Gamma = \left[\int_{r_c-x_D}^{r_c+x_D}
  [rP+r(\Omega-\Omega_p)\Sigma j]dr\right]_F^R,
\end{equation}
where the $R$ superscript indicates that the integral has to be
performed on the rear side of the domain $D$ (radial boundary at $\phi
=-\phi_D< 0$), while the $F$ subscript indicates that the integral has
to be performed on the front side of the domain $D$ (radial boundary
at $\phi=\phi_D > 0$). In equation~(\ref{eq:5}), the first term of the
integrand represents the pressure torque exerted on the material
enclosed within the domain~$D$, while the second term represents the
budget of angular momentum brought to this region by advection. Since
the flow is steady in the corotating frame, the angular momentum of
this domain is constant in time and the torque is therefore integrally
transmitted to the planet.  We recognize in the second term of the
integrand of equation~(\ref{eq:5}) the classical horseshoe drag expression
\citep{wlpi91,wlpi92,masset01,mp03,mak2006,2009arXiv0901.2265P}, but
this equation also shows the pressure contribution, which has been
overlooked in previous analysis.

\subsection{A Bernoulli invariant}
\label{sec:BernDef}
We derive, from equations~(\ref{eq:eulerrad}) and~(\ref{eq:euleraz})
in steady state, a Bernoulli invariant, constant along streamlines in
steady state:
\begin{equation}
\label{eq:6}
B = \frac{v^2}{2} + \eta + \Phi - \frac{r^2 \Omega_p^2}{2},
\end{equation}
where $v^2=v_r^2+v_\phi^2$.

Far from the planet (i.e. in $\phi_D$ and $-\phi_D$), the streamlines
are purely azimuthal, and the fluid elements do not feel the planet's
potential, so we get, neglecting the indirect term:
\begin{equation}
\label{eq:bernfar}
B = \frac{1}{2}r^2\left(\Omega - \Omega_p \right)^2 + 
\phi_\ast(r) - \frac{1}{2} r^2\Omega_p^2 + \eta
\end{equation}
Following \cite{mp03}, we derive:
\begin{equation}
\label{eq:berndr}
\partial_r B = r \omega \left(\Omega - \Omega_p \right),
\end{equation}
where we used the radial equilibrium: $\partial_r\phi_\ast =
r\Omega^2-\partial_r\eta$.

Before going further and inject this into equation~(\ref{eq:5}), we evaluate
how the Bernoulli invariant is modified at a given location in the
disk with respect to its value in the unperturbed disk.  Considering a
small perturbation $\delta r$ of radius, $\delta \eta$ of enthalpy,
and $\delta \Omega$ of angular velocity, the total variation of the
Bernoulli invariant is:
\begin{eqnarray}
\label{eq:7}
\delta B &=&  r \omega \left(\Omega - \Omega_p \right) 
\delta r -\partial_r \eta\ \delta r+ \delta \eta \\
\label{eq:bernpert}
&=& \left(\Omega - \Omega_p \right) \delta j + \delta' \eta,
\end{eqnarray}
where $\delta' \eta = \delta \eta -\partial_r \eta\ \delta r $ is the
\emph{local} perturbation of enthalpy, at fixed radius.

\subsection{Final torque expression}
We now evaluate the total torque, using equations~(\ref{eq:5}) and
(\ref{eq:berndr}), which allows to change the variable of integration
to $B$. We use the subscripts $(1)$ to $(4)$ to refer to the quadrants 
of figure \ref{fig:hsregion}, and we get:
\begin{eqnarray}
\label{eq:8}
\Gamma&=&\left[\int_{r_c-x_D}^{r_c+x_D}
  [rP+r(\Omega-\Omega_p)\Sigma j]dr\right]_F^R\\
&=& \int_{B_\infty\ (1)}^{B_c} \frac{\Sigma}{\omega} j dB   
+   \int_{B_c\ (2)}^{B_\infty}  \frac{\Sigma}{\omega} j dB \\\nonumber
&& - \int_{B_\infty\ (3)}^{B_c}  \frac{\Sigma}{\omega} j dB   
-  \int_{B_c\ (4)}^{B_\infty}  \frac{\Sigma}{\omega} j dB \\\nonumber
&&+  \int_{r_c-x_D\ (1)(2)}^{r_c+x_D} rP \ dr -
\int_{r_c-x_D \ (3)(4)}^{r_c+x_D} rP \ dr,
\end{eqnarray}
where we supposed the radial boundaries to be far enough in $r_c\pm
x_D$ so we can assume the streamlines to be purely circular there,
corresponding to a Bernoulli invariant $B_\infty$.

We note with a subscript $0$ the values of the unperturbed flow.
The perturbation of angular momentum and pressure are
denoted $\delta j$ and $\delta P$, and we use the conservation of
vortensity along a horseshoe U-turn in order to write
$\left. \Sigma/\omega \right|_{1(4)}=\left. \Sigma/\omega
\right|_{2(3)}$. This leads to:

\begin{eqnarray}
\label{eq:9}
\Gamma
&=& \int_{B_\infty\ (1)}^{B_c} \left. \frac{\Sigma}{\omega}\right|_1 (j_0^- +\delta j^R_1) dB   
-   \int_{B_\infty\ (2)}^{B_c}  \left. \frac{\Sigma}{\omega} \right|_1 (j_0^+ +\delta j^R_2) dB \\\nonumber
&& - \int_{B_\infty\ (3)}^{B_c} \left.  \frac{\Sigma}{\omega} \right|_4 (j_0^- +\delta j^F_3) dB
+  \int_{B_\infty\ (4)}^{B_c}  \left. \frac{\Sigma}{\omega} \right|_4 (j_0^+ +\delta j^F_4) dB \\\nonumber
&&+  \int_{r_c-x_D\ (1)}^{r_c+x_D} r(P_0^- +\delta P^R) \ dr +  \int_{r_c-x_D\ (2)}^{r_c+x_D} r(P_0^+ +\delta P^R) \ dr \\\nonumber
&&-  \int_{r_c-x_D\ (3)}^{r_c+x_D} r(P_0^- +\delta P^F) \ dr -  \int_{r_c-x_D \ (4)}^{r_c+x_D} r(P_0^+ +\delta P^F) \ dr,
\end{eqnarray}

The integrals over $rP_0^\pm$ simplify, between (1) and (3) on the one
hand, and (2) and (4) on the other hand.  We denote $\Delta j_0$ the
total jump of angular momentum of a fluid element between the inside
and outside region: $\Delta j_0(B) = j^0_+(B) - j^0_-(B)$, assuming
that before and after the jump, the fluid element has the specific
angular momentum of the {\em unperturbed} flow corresponding to its
value of the Bernoulli invariant. By construction $\Delta j_0(B)$ is
always a positive quantity.

Using equation~(\ref{eq:bernpert}), we obtain:
\begin{equation}
\label{eq:dj_f_deta}
\delta j^R =-\frac{\delta' \eta^R}{ \left(\Omega - \Omega_p \right)}.
\end{equation}
Since we have $\delta ' \eta =\delta P/\Sigma$, and $dr
= \partial_{B}r dB = dB/(r\omega(\Omega-\Omega_{p}))$, we also can
recast the integrals over perturbed pressure using $B$ as an
integration variable. We are hence left with:
\begin{eqnarray}
\Gamma
&=& - \int_{B_\infty}^{B_c} \left. 
  \frac{\Sigma}{\omega}\right|_1 \Delta j_0(B) dB 
-   \int_{B_\infty  (1)}^{B_c} \left. 
  \frac{\Sigma}{\omega} \right|_1 \frac{ \delta' \eta^R}{ \left(\Omega - \Omega_p \right)} dB 
+   \int_{B_\infty  (2)}^{B_c} \left. 
  \frac{\Sigma}{\omega} \right|_1 \frac{\delta' \eta^R}{ \left(\Omega - \Omega_p \right)} dB \nonumber\\
&& + \int_{B_\infty}^{B_c} \left.  
  \frac{\Sigma}{\omega} \right|_4 \Delta j_0(B) dB
+  \int_{B_\infty  (3)}^{B_c}  \left. 
  \frac{\Sigma}{\omega} \right|_4 \frac{\delta' \eta^F}{ \left(\Omega - \Omega_p \right)} dB
-  \int_{B_\infty  (4)}^{B_c}  \left. 
  \frac{\Sigma}{\omega} \right|_4 \frac{\delta' \eta^F}{ \left(\Omega - \Omega_p \right)} dB \nonumber\\
&&
+  \int_{B_\infty \ (1)}^{B_c} r \delta ' 
\eta^R\Sigma \ \frac{dB}{r\omega(\Omega-\Omega_{p})}  
-  \int_{B_\infty \ (2)}^{B_c} r \delta ' 
\eta^R\Sigma \ \frac{dB}{r\omega(\Omega-\Omega_{p})}  
\nonumber\\
\label{eq:10}
&&-  \int_{B_\infty \  (3)}^{B_c} r \delta ' \eta^F\Sigma  
\frac{dB}{r\omega(\Omega-\Omega_{p})}
+  \int_{B_\infty \ (4)}^{B_c} r \delta ' \eta^F\Sigma  
\frac{dB}{r\omega(\Omega-\Omega_{p})}.
\end{eqnarray}
If we denote with $B_s$ the value of the Bernoulli invariant of the
separatrices, we immediately see that $\Delta j_0$ is strictly equal
to $0$ for $B_{\infty}<B<B_{s}$ (i.e. in the region that is not part
of the horseshoe region, so that its particles do not undergo any
jump), so the integrals over $\Delta j_0$ reduces to their values
between $B_s$ and $B_c$.  Another simplification arises from the fact
that the integrals which were previously over $\delta P$ --~i.e. the
last four terms of equation~(\ref{eq:10})~-- and over $\delta j$
--~the second, third, fifth and sixth terms of
equation~(\ref{eq:10})~-- cancel each other, and we are only left
with, reverting the integral boundaries:
\begin{eqnarray}
\label{eq:TorqueFinal}
\Gamma
&=& \int_{B_c}^{B_s} \left. \frac{\Sigma}{\omega}\right|_1 \Delta j_0(B) dB  -
\int_{B_c}^{B_s} \left.  \frac{\Sigma}{\omega} \right|_4 \Delta j_0(B) dB ,
\end{eqnarray}
which is exactly the original expression derived by \cite{wlpi91}.
This derivation shows that the whole contribution of the material in
the vicinity of corotation, including the material that does not
belong to the horseshoe region and circulates beyond the separatrices,
reduces to a standard horseshoe drag integral that only involves the
vortensity profile inside of the horseshoe region.

\subsection{A model of pressure waves}
\label{sec:pressure_model}

Although the torque is of primary interest, it is also worthwhile to
have a peek at the anatomy of the horseshoe region, and to stress its
major properties. Our first step consists in evaluating the pressure
response to the horseshoe dynamics, azimuthally far from the planet.

Before going further, we need to know the response of the disk to any
small arbitrary perturbation of vortensity, independent of the
azimuth.  Indeed, the horseshoe dynamics creates, in the downstream
sides of the horseshoe region, stripes of perturbed vortensity, which
trigger a density (or pressure) and velocity response. We denote these
quantities respectively with $\delta \Sigma$ and $\delta v_\phi$.  We
start with the perturbed rotational equilibrium (equation
\ref{eq:eulerrad}):
\begin{equation}
\label{eq:11}
-2\Omega_à\delta v_\phi+c_s^2 \left(\frac{\partial_x\delta \Sigma}{\Sigma_c}
+\frac{\partial_x \Sigma_c}{\Sigma_c}\frac{\delta\Sigma}{\Sigma_c}\right)=0 ,
\end{equation}
where the pressure is
$P=c_s^2\Sigma$. Deriving this equation with respect to $x$ leads to:
\begin{equation}
\label{eq:12}
-2\Omega_p\partial_x \delta v_{\phi} + c_s^2 \frac{\partial^2_x\delta\Sigma}{\Sigma_c}=0,
\end{equation}
where, for a given quantity $\xi$ we neglected all the terms in
$\partial_x \xi$ comparatively to the ones in $\partial_x \delta \xi$,
since, as we shall see, the latter scale with $1/H$, which is much
larger than the $1/r$ scaling of the former.  In order to eliminate
the unknown $\delta v_\phi$ in this equation, we make use of the
perturbed vortensity:
\begin{equation}
\label{eq:13}
\delta w = \frac{\partial_x \delta v_\phi}{\Sigma_c} - 
\frac{\omega_0\delta\Sigma}{\Sigma_c^{\ 2}},
\end{equation}
so we can write:
\begin{equation}
\label{eq:70J}
\delta\Sigma - \frac{c_s^2}{\kappa^2} \partial_x^2 
\delta \Sigma = -\Sigma_c \frac{\delta w}{w_0},
\end{equation}
where $\kappa = (2\Omega_p\omega_c)^{1/2}$ is the epicyclic frequency.

The general solution of Eq.~(\ref{eq:70J}) is the convolution product
of its right hand side by the Green's kernel $K(x)$, which is the
solution of:
\begin{equation}
  \label{eq:14}
  \delta \Sigma-\frac{c_s^2}{\kappa^2}\partial_{x}^2\delta \Sigma = \delta(x),
\end{equation}
and whose expression is:
\begin{equation}
  \label{eq:15}
  K(x) = \frac{1}{2H}e^{-|x|/H},
\end{equation}
where we have specialized to the Keplerian case, for which $H\equiv
c_s/\Omega=c_s/\kappa$.  This kernel represents the pressure response
to a singular perturbation of $w$ at $x=0$, of weight $\int
w(x)dx=w_0/\Sigma_c$. We note that $K(x)$ has a unitary weight:
\begin{equation}
  \label{eq:16}
  \int_{-\infty}^{+\infty}K(x)dx = 1.
\end{equation}
Downstream of the flow, we obtain $\delta w$ straightforwardly by
making use of the conservation of vortensity along a streamline, and
assuming radially symmetric U-turns (so that fluid particle undergoing
a jump at $r=r_c+x$ is mapped to $r=r_c-x$). For the part behind the
planet:
\begin{eqnarray}
\label{eq:17}
\delta w^R(x) &=& -2\frac{x}{r_c}w_0 \mathcal{V} \
\textrm{for} \ 0 < x < x_s\\\nonumber
&=&0\  \textrm{elsewhere}.
\end{eqnarray}
Similarly, for the part in front of the planet
\begin{eqnarray}
\label{eq:18}
\delta w^F(x) &=& -2\frac{x}{r_c} w_0\mathcal{V} \ 
\textrm{for} \ 0 > x > -x_s\\\nonumber
&=&0\  \textrm{elsewhere}.
\end{eqnarray}
If we assume $x_s \ll H$, this gives for the perturbed surface density
(at the rear of the planet, for the sake of definiteness):
\begin{eqnarray}
\label{eq:19}
\delta \Sigma^R (x) &=&K \ast \Sigma_c \frac{\delta w}{w_0}\\
\label{eq:20}
&=&K \ast \left(-\frac{2x}{r_c}\mathcal{V}\Sigma_c\right)\\
\nonumber
&\simeq& \int_0^{x_s} \frac{1}{2H} \frac{2x\mathcal{V}}{r_c}\Sigma_cdx\\
\label{eq:21}
\delta \Sigma^R (x) &=& \Sigma_c\frac{x_s^2\ \mathcal{V}}{2H r_c} 
\end{eqnarray}

The equation~(\ref{eq:20}) shows that the actual profile is obtained
by convolution of the fictitious surface density profile
$(-2x/r_c)\mathcal{V}\Sigma_c$ with the kernel $K$ of unitary weight,
which amounts to spreading radially this fictitious profile without
changing the total mass of the perturbation. One can check that this
fictitious profile is the profile that one would obtain by attributing
the vortensity perturbation to the perturbed surface density, without
altering the rotation profile (i.e. keeping the profile of vorticity
of a Keplerian disk). This helps understanding why the horseshoe drag
in a pressure supported disk does not differ from the horseshoe drag
estimated by considering test particles \citep{wlpi91}.

We immediately deduce the corresponding perturbation of enthalpy
(which is therefore flat over the horseshoe region because of our
hypothesis that $x_s\ll H$):
\begin{equation}
\label{eq:d_etaR}
\delta \eta^R = T_c \frac{x_s^2\ \mathcal{V}}{2H r_c}
\end{equation}
Similarly, we have:
\begin{equation}
\label{eq:d_etaF}
\delta \eta^F = -T_c \frac{x_s^2\ \mathcal{V}}{2H r_c}
\end{equation}

\subsection{Impact on the horseshoe width}
\label{sec:width} 
The value of the Bernoulli invariant at the separatrices is the same
as its value at the stagnation point. Quite in contrast with the
situation of a non-barotropic disk, for which we shall see in paper~II
that the value of the invariant is discontinuous at the stagnation
point, here the value of the Bernoulli invariant is necessarily
continuous in the vicinity of the stagnation point, and the
separatrices of the horseshoe region all share the same value $B_s$ of the
Bernoulli invariant.  As a consequence, in order
to ensure that it is associated to a Bernoulli invariant with value
$B_s$, a given separatrix must shift radially, by a value $\delta
x_s$, in order to compensate for the variation of the enthalpy due to
the evanescent waves. At the rear of the planet, we have, using
equations (\ref{eq:d_etaR}) and (\ref{eq:bernpert}), evaluated at
$x=x_s$:

\begin{eqnarray}
\label{eq:22}
\delta B = 0 = 4A_pB_px_s \delta x_s^R +T_c \frac{x_s^2\ \mathcal{V}}{2H r_c} \\
\frac{\delta x_s^R}{x_s} = - \frac{T_c}{8AB r_c}\frac{ \mathcal{V}}{H}
\end{eqnarray}
Or, specializing to the case of a Keplerian disk: 
\begin{equation}
\label{eq:dxsr}
\frac{\delta x_s^R}{x_s} = - \frac{2}{3} \mathcal{V}h.
\end{equation}
A similar result holds in front of the planet: 
\begin{equation}
\label{eq:dxsf}
\frac{\delta x_s^{F}}{x_s} = \frac{2}{3} \mathcal{V}h.
\end{equation}
We deduce the estimate of the relative asymmetry of the horseshoe region:
\begin{equation}
  \label{eq:23}
  \frac{x_s^F-x_s^R}{(x_s^F+x_s^R)/2}=\frac 43\mathcal{V}h
\end{equation}

The horseshoe region therefore becomes asymmetric when there is a
vortensity gradient. The relative asymmetry scales with the vortensity
gradient and the disk's aspect ratio, and it is independent of the
planetary mass.

\section{Numerical simulations}
\label{sec:Numeric}
We performed simulations of our system using the FARGO code
\citep{fargo2000,fargo2000b}, in order to check the properties
predicted by the above analysis.
\subsection{Numerical Setup}
\label{sec:numsetup}
All our runs are performed starting with a reference run, and then
changing only one parameter. The spatial resolutions of these runs is
$700$~zones in azimuth and $474$~in radius, for a ring spawning
radially from $0.8 a$ to $1.2 a$. We stress that, even if the
radial width of our mesh seems small, we apply damping conditions at
the boundaries that ensure no reflection of the wake. The damping
factor was tuned in order to ensure that this assumption is valid.

We use a planet with a mass ratio to the primary of $8 \times 10^{-6}$
(approximately $2.6$~$M_\earth$, for a central star of mass
$1$~$M_\sun$).  The initial conditions are presented in section
\ref{sec:notations}. For the surface density, we use a typical value
of $\sigma=-1$ (this value is not realistic as it corresponds to a
surface density increasing outwards, but it corresponds to a large
vortensity gradient which exacerbates the effects that we wish to
check). The surface density at corotation is $\Sigma_c = 2 \times
10^{-3}$, and the aspect ratio $h$ is set to 0.05 at $r=r_c$.

The last important parameter is the potential softening length.  We
take it to be $30$\% of the disk thickness, which is a rather low
value.  Changing this value can have significant effects, which are
discussed in section \ref{sec:softening}.

We indeed observe that the horseshoe region is asymmetric, by
comparing its front and rear width. We measure it at $\phi=\pm
1$~rad., far enough not to feel the planet's potential, and once the
steady state is reached (after about $25$~orbits). As it is taken in
steady state, when the separatrix is well defined, we measure the
whole full width (hereafter FW) of the separatrix, rather than twice the
upstream half-width (distance between the upstream separatrix and the
corotation, hereafter TUHW). The difference between the two methods
turned out to be about 3-4\%, which is small, but about the same order
than the effect we are looking for. In fact, we can see this as an
effect of pressure. Here, contrary to our assumptions of section
\ref{sec:pressure_model}, we do not strictly have $x_s \ll H$ ($x_s
\sim 0.01a$), so the shape of the pressure wave plays a small role,
with a larger amplitude on the downstream part of the U-turn. To get
rid of this problem, we chose to use the FW, more appropriate here.

\subsection{Dependence on the vortensity gradient}
\label{sec:vortensity_grad}
\begin{figure}
	\centering
		\plotone{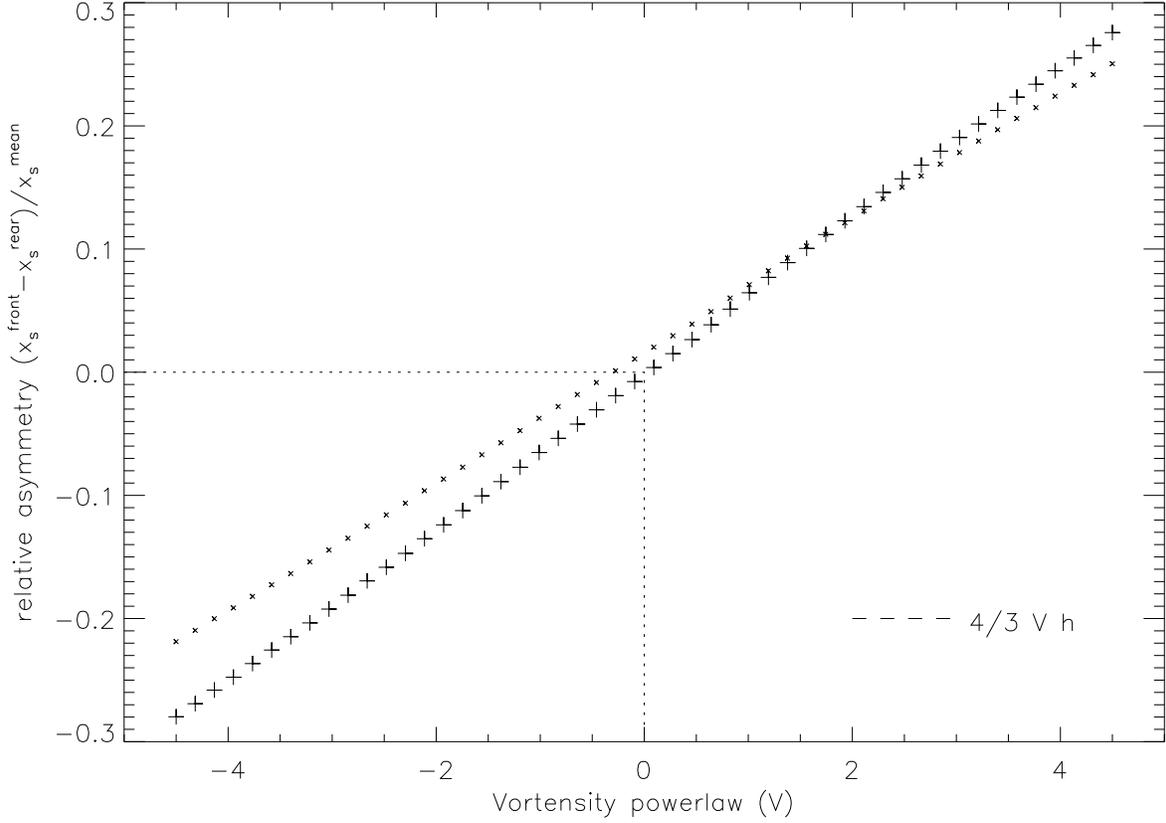}
                \caption{Relative asymmetry as a function of the
                  vortensity gradient.  The small and large crosses
                  are extracted from simulations, the large crosses representing 
                  the asymmetry resulting from the measurement of the FHW, while 
                  the small ones correspond to the TUHW (see text). The dashed
                  line represents our analytical expression of
                  equation~(\ref{eq:23}). The fit is very
                  satisfactory, except at very large gradients.  Note
                  the relatively high values of the asymmetry: for
                  realistic values of $\mathcal{V}$, the difference in
                  size between the front and rear region can amount to
                  $15$~\% of the ``unperturbed horseshoe width''
                  (which is taken to be the mean of the front and rear
                  widths; see section \ref{sec:defHS}).}
	\label{fig:xsfv}
\end{figure}

Our first step was to conduct simulations with a varying vortensity
gradient, in order to quantify the front-rear asymmetry of the
horseshoe region. We performed $50$ runs with a value of ${\cal V}$
varying from $-4.5$ to $4.5$. The results are presented in figure
\ref{fig:xsfv}. As we measure the relative asymmetry, the discrepancy
between the FW an the TUHW is obvious here, the effect being cumulative
(for a positive $\mathcal{V}$, the front FW is greater than the TUHW,
while the rear FW is lower than the TUHW). The fit of the numerical
results with our expression is very good.

\subsection{Dependence on the aspect ratio}
\label{sec:aspect_ratio}
\begin{figure}
	\centering
		\plotone{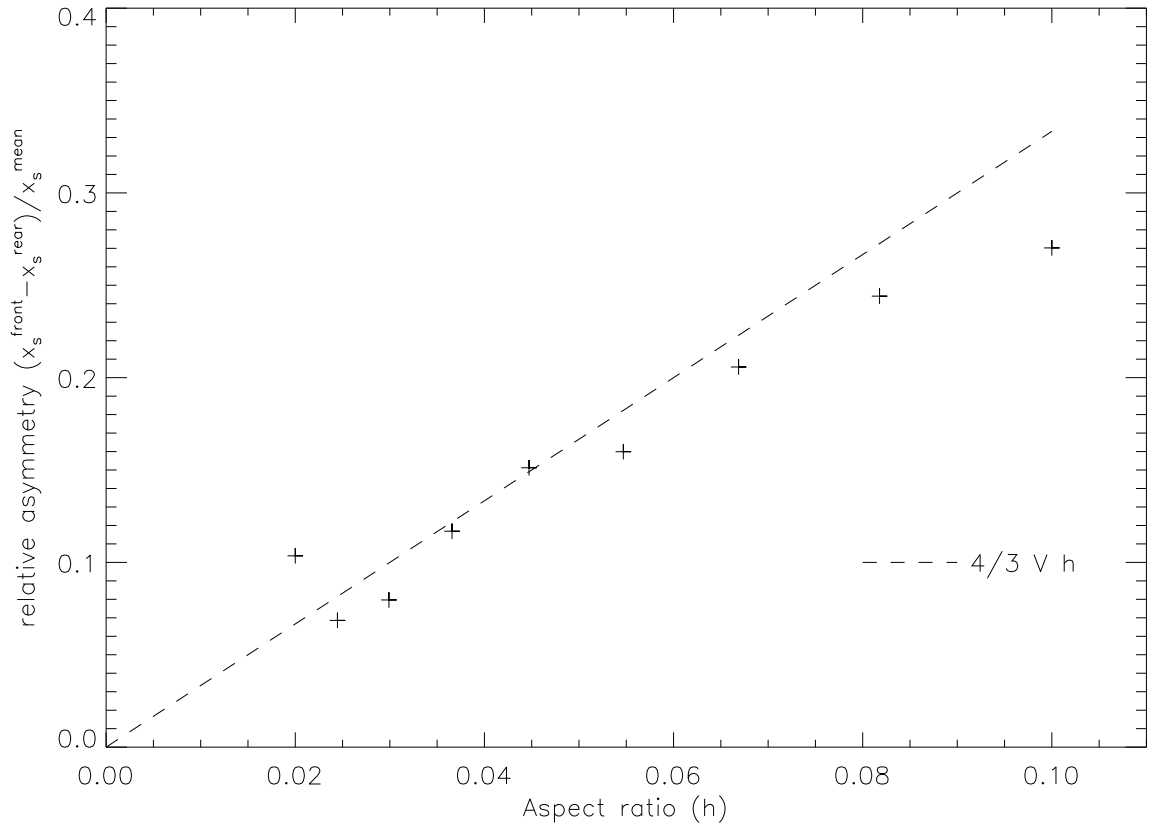}
                \caption{Relative asymmetry as a function of the
                  aspect ratio. The dashed line represents our
                  analytical expression. The points are extracted from
                  simulations, which have the standard parameters,
                  except for the aspect ratio. Again, there is a
                  satisfactory agreement, albeit not as good as the
                  dependency on $\mathcal{V}$.}
	\label{fig:xsfh}
\end{figure}

The other factor impacting equation (\ref{eq:23}) is the aspect
ratio. We checked this dependency for some values of $h$, as shown in
figure \ref{fig:xsfh}. The agreement with the theory is good, but not
perfect.  It is presumably related to modifications in the topology of
the flow, which induce small discrepancies between the runs. This was
not a problem for the runs with a varying $\mathcal{V}$, where the
streamline topology remained essentially unchanged between the
different runs.

\subsection{Pressure profile}
\label{sec:pressure_profile}

\begin{figure}
	\centering
		\plotone{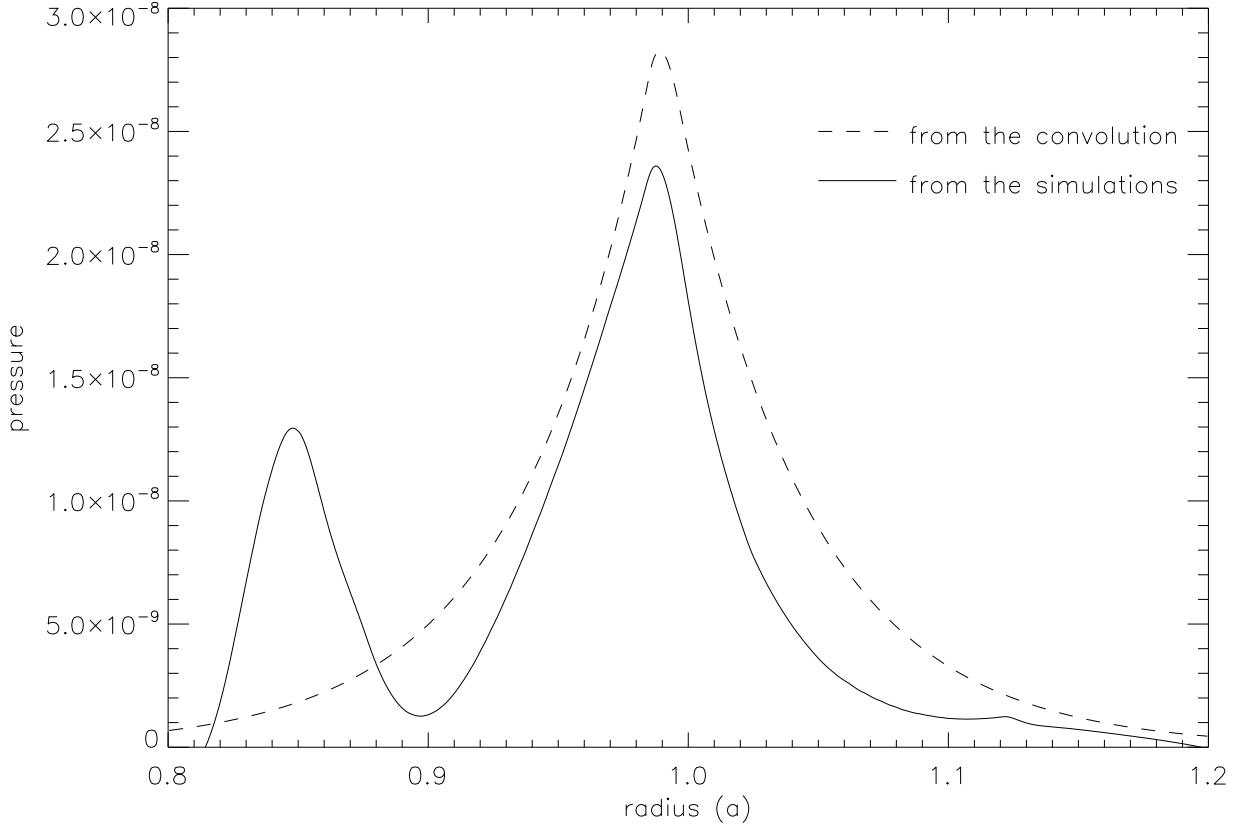}
                \caption{Radial pressure profile. This plot is
                  obtained at an azimuth of $\phi=1$, for a run with
                  the standard parameters of section
                  \ref{sec:numsetup}. We compare the theoretical
                  profile obtained by convolution of the vortensity
                  perturbation by the evanescent waves kernel,
                  properly weighted, as indicated by
                  equation~(\ref{eq:19}), and the outputs from the
                  simulations. The shapes are very similar in the
                  corotation region (our model, which assumes that the
                  response is taken sufficiently far from the planet
                  to be invariant in $\phi$, breaks down in the region
                  of Lindblad resonances because of the presence of
                  the wake).}
	\label{fig:pfr}
\end{figure}

In section \ref{sec:pressure_model}, we considered $x_s \ll H$, so
that the perturbed surface density, spread over a distance $H$, is
uniform across the horseshoe region. As we have seen, this assumption
is not always true, and we try here to relax this assumption, modeling
the perturbed surface density as a convolution product of the
perturbed vortensity by the Green function $K(x)$, properly
normalized. We compare this to the perturbed surface density of our
simulations in figure \ref{fig:pfr}. We stress that we have used, for
$\delta w$, a truncated triangular profile, in order not to take into
account the material close to corotation that did not have time to
perform a U-turn. Our simulated profile agrees well with the measured
profile from the simulations, both in amplitude and position. The
differences on amplitude may be imputed to an uncertainty in $x_s$,
which is not perfectly defined here, as the situation is not {\em
  stricto sensu} in steady state.

\subsection{Time evolution}
\label{sec:time}
\begin{figure}
	\centering
		\plotone{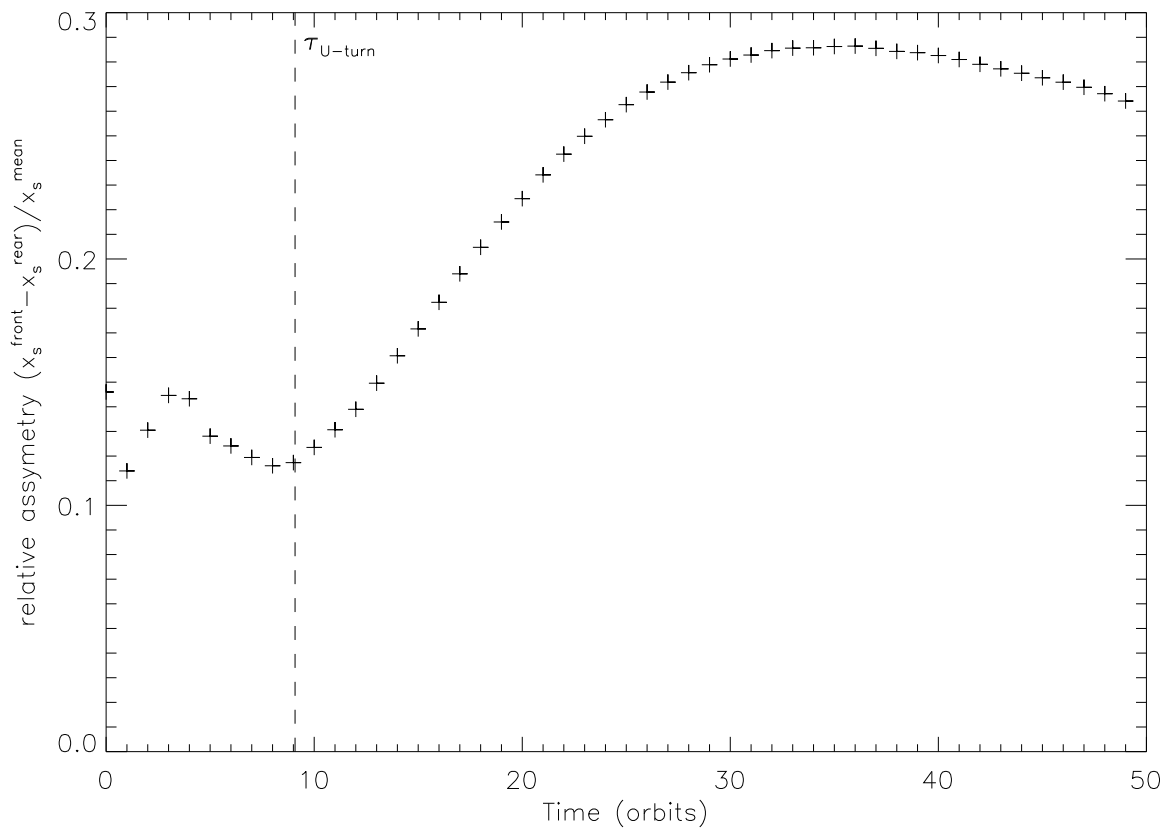}
                \caption{Relative asymmetry as a function of time, in
                  orbits. This is a run with standard parameters,
                  except that the aspect ratio is $h=0.08$. The
                  vertical dashed line represent
                  $\tau_{U-turn}$. Here, the asymmetry boost starts
                  around $\tau_{U-turn}$, as expected. Note that the
                  initial asymmetry is not $0$, presumably due to the
                  initial pressure gradient in the disk (disks with
                  vanishing pressure gradients do not exhibit initial
                  asymmetry).}
	\label{fig:xsft}
\end{figure}

The asymmetry of the horseshoe zone takes some time to establish. When
the planet is ``turned on'' in the disk, the upstream region remains
unperturbed for a while. A typical time for pressure waves to develop
is the time needed for particles to perform a horseshoe U-turn.
Following \citet{bm08}, this time is approximatively:
$\tau_{U-turn}=\Omega_p^2h^{3/2}q^{-1/2}/(A_p^2B_p)$. This is in good
agreement with our simulations, as shown in figure
\ref{fig:xsft}. Here, a small initial asymmetry (approx. $1/3$ of the
final asymmetry) already exists, which we did not investigate, owing to
its transient behavior.

\subsection{Impact of the mass}
\label{sec:mass}
\begin{figure}
	\centering
		\plotone{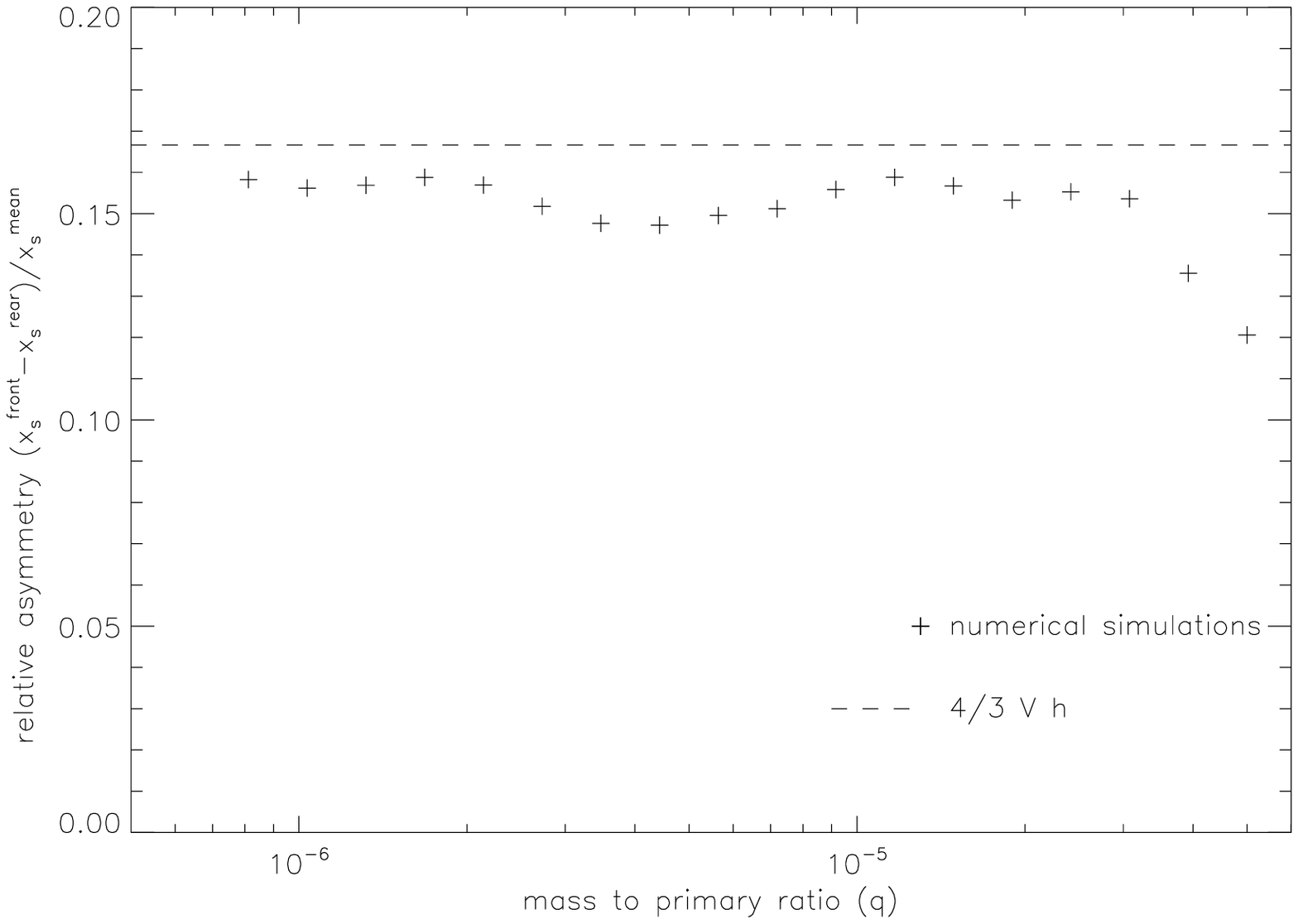}
                \caption{Relative asymmetry as a function of the
                  planet mass to primary mass ratio.  The dashed line
                  represents our analytical expression, given by
                  equation~(\ref{eq:23})). The other points are
                  extracted from simulations. Note that each point is
                  measured at a different date, $t_{\rm
                    measure}\propto 1/\sqrt{q}$, as the locally steady
                  state takes more time to reach for smaller planets.}
	\label{fig:xsfq}
\end{figure}

Since we considered the low-mass planet limit, we want to know to
which extent our expression is true. To do so, we performed
calculations varying the planet mass to primary mass ratio (see figure
\ref{fig:xsfq}). We see that our expression stand up to $q \simeq
2\times 10^{-5}$ (approx. $7$~$M_\earth$), significantly over the
usual ``low-mass'' planet limit \citep{mak2006}. Presumably this is
because the condition $x_s<H$ can be marginally verified while
equations~(\ref{eq:dxsr}) and~(\ref{eq:dxsf}) still hold.

\section{Discussion}
\label{sec:discussion}

We discuss hereafter a number of points related to our results.

\subsection{A definition of the horseshoe width}
\label{sec:defHS}
The torque expression of equation~(\ref{eq:TorqueFinal}) is the one
derived by \citet{wlpi91}, except that the integral boundaries are
expressed in terms of values of the Bernoulli invariant, rather than
$x_s$. For practical purposes, one would wish to have an expression in
terms of the horseshoe half width.  The latter is ill-defined,
however, owing to the asymmetry of the horseshoe zone.  In order to
provide an adequate definition of the average horseshoe width, we need to derive $B_s$, getting rid of $\delta \eta '$. To achieve this, although they share the same value $B_s$, we explicitly sum $B^F_s$ and $B^R_s$, which yields, to second order in $x$:
\begin{eqnarray}
\label{eq:24}\nonumber
B^F_s + B^R_s &=& B_c + 2A_p B_p(x^F_s)^2 + \delta' 
\eta^F + B_c+ 2A_p B_p(x^R_s)^2 + \delta' \eta^R  \\
B_s&=&B_c + 2A_p B_p\ \frac{(x^F_s)^2+(x^R_s)^2}{2}
\end{eqnarray}
In order to transform equation~(\ref{eq:TorqueFinal}) so as to get rid
of the Bernoulli invariant, we write to the first order in $x$ the
factors of the integrand: $\Delta j_0(x)=4B_p r_c x$ and
$\Sigma/\omega=\Sigma_c/\omega_c(1+\mathcal{V} x/r_c$). This yields:
\begin{eqnarray}
\label{eq:25}\nonumber
\Gamma &=& 4B_p r_c \frac{\Sigma_c}{\omega_c}
\left[\int_{B_c}^{B_s} |x| 
  \left(1-{\mathcal{V}}\frac{|x|}{r_c}\right) dB - 
  \int_{B_c}^{B_s} |x| 
  \left(1+\mathcal{V}\frac{|x|}{r_c}\right)dB\right] \\\nonumber
&=&-8B_p \frac{\Sigma_c}{2B_p} \mathcal{V} 
\int_{B_c}^{B_s}\frac{B-B_c}{2A_pB_p}dB \\
&=&\frac{3\Omega^2}{4}\Sigma_c\mathcal{V}
\left[\frac{(x^F_s)^2+(x^R_s)^2}{2}\right]^2,
\end{eqnarray}
which is the same torque expression as the one of \citet{wlpi91},
provided we take:
\begin{equation}
\label{eq:xmean}
x_{s}^{\rm mean} = \left[\frac{(x^F_s)^2+(x^R_s)^2}{2}\right]^{1/2}
\end{equation}
This gives an adequate definition of the horseshoe width to be used
when one wishes to estimate the horseshoe drag by means of a
streamline analysis.

\subsection{A direct estimate of the corotation torque} \label{simplified}
As we have seen, taking into account the pressure effects does not
unveil a different torque value from the one derived by
\citet{wlpi91}. The main reason that we have mentioned in
section~\ref{sec:pressure_model} is the fact that the launch of
evanescent waves in the coorbital region renders radially more diffuse
the perturbation of surface density (which would otherwise be bound to
the horseshoe region), but leaves its linear density unchanged. This result
is not surprising, since the fluid model of \citet{gt79} had no pressure dependence.

One can wonder, however, why the horseshoe region asymmetry described
in section~\ref{sec:width} does not have any impact on the torque. In
order to gain some insight into the reasons for that, we examine here
how this asymmetry alters the front and rear components of the torque
by a direct, albeit approximate, estimate of the corotation torque.

The horseshoe U-turns, by conserving the vortensity, produce stripes
of perturbed vortensity in front of the planet and behind it,
in a disk with a vortensity gradient. These perturbation of vortensity
yield perturbations of surface density. Disregarding the radial spread
of these perturbations of surface density, we can write their linear
mass as:
\begin{equation}  
\Lambda = \int \delta\Sigma(x)dx =\frac{\mathcal{V}x_s^2\Sigma_c}{a}
\end{equation}  
in front of the planet, and:
\begin{equation}  
\Lambda =-\frac{\mathcal{V}x_s^2\Sigma_c}{a}
\end{equation}  
behind the planet.  We evaluate the torque arising from a stripe of
linear mass $\Lambda$ onto the planet, assuming that it starts at a
distance $d$ from the latter:
\begin{equation}
\Gamma_{\rm stripe}  \approx  a\int_d^\infty r_c 
\frac{GM_p}{l^2}\Lambda dl = \frac{\Omega_p ^2a^4q}{d}\Lambda.
\label{eq:GammaStrip}
\end{equation}
We note that the effects of the stripe in front of the planet, and of
the stripe that is located behind it, are cumulative, since they
involve perturbations of surface density of opposite signs, hence the
corotation torque, in total, is:
\begin{equation}
  \label{eq:26}
  \Gamma \approx \Omega_p^2a^3q\mathcal{V}
  \Sigma_c\frac{(x_s^F)^2+(x_s^R)^2}{d}.
\end{equation}
This allows to understand why the horseshoe drag remains unchanged
when the horseshoe region becomes asymmetric: $x_s^F$ increases
(decreases) while $x_s^R$ decreases (increases), but their sum (or the
sum of their square) remains constant to lower order in $\mathcal{V}$.
Put into simple words, taking for example a positive torque excess
($\mathcal{V}>0$), the widening of the front region produces a
supplementary torque excess, since the front region is producing a
positive torque excess.  But the torque excess produced by the rear
region is also positive, so the shrinking of this region induces a
smaller torque excess, which balance, to the first order, the
supplementary torque excess of the front region, since $|\delta
x_s^F|=|\delta x_s^R|$.

As a side result, we compare equation~(\ref{eq:26}) to the expression
of \citet{wlpi91}, which reads with our notation, in a Keplerian disk:
\begin{equation}
\Gamma_{\rm Ward} = \frac{3}{4} \mathcal{V} \Sigma_c x_s^4 \Omega_p^2.
\end{equation}
Noting that $x_s^2\sim a^2q/h$ \citep{mak2006,2009arXiv0901.2263P},
and assuming that the stripes originate at a distance $\sim H$ from
the planet, equation~(\ref{eq:26}) can be recast as:
\begin{equation} 
\Gamma \sim 2\mathcal{V} \Sigma_c x_s^4 \Omega_p^2
\end{equation} 
This is the same expression except for a $3/8$ factor, presumably
originating from our simple estimates of $d$ and $x_s$.

\section{Relaxation of the globally isothermal disk hypothesis}
In a globally isothermal disk, the asymmetry of the horseshoe region
does not involve additional generation of perturbed surface density,
because there is no generation of vortensity, since the gradients of
pressure and density are always aligned.  However, in a locally
isothermal disk, where the temperature depends on the radius ($T=T(r)
= T_c\ (r/r_c)^{-\tau}$), this is not true anymore, and the vortensity
created corresponds to an additional perturbation of surface density
that exerts a torque onto the planet.  For the sake of definiteness we
call this torque the non-isothermal torque excess.  For a proper
formulation of this torque, one could follow the method of paper~II,
using an adequate invariant along the streamlines, whenever it exists,
in order to derive the downstream vortensity and perturbed surface
density.  In the present situation, nevertheless, the variation of the
state variables of a fluid element that goes from infinity to the
stagnation point depends on its path, so we cannot use the method of
paper~II for adiabatic flows.  We expect therefore the non-isothermal
torque excess to display a strong sensitivity to the topology of the
streamlines. As we shall see hereafter, this is indeed the case.

Prior to a fine tuned description of the excess inferred from
numerical simulations, we give hereafter an oversimple estimate of the
magnitude of the offset, which should apply to situations in which
there is only one stagnation point, and which is adapted from a
discussion provided in paper~II, in which we interpret the adiabatic
torque excess as arising from a constitutive asymmetry of the
horseshoe region.

\subsection{An oversimple approach}
\label{sec:model_LocIsoT} 
We know from our simulations that a locally isothermal disk exhibits a
non-isothermal torque excess, even for a vanishing vortensity
gradient.  In order to estimate this excess, we provide hereafter an
estimate of the intrinsic asymmetry of the horseshoe region of a
locally isothermal disk. All this study is done for $\mathcal{V}=0$,
in order to neglect the asymmetry of the horseshoe region acquired on
the long term under the feed back of the evanescent waves, which does
not plays a role in the torque, as explained before.

To estimate the intrinsic asymmetry, we consider a fluid element
belonging to a separatrix, that starts at $(x_s, \phi_\infty)$. By
definition the torque applied to it is exactly sufficient to bring it
to the stagnation point. For the globally isothermal case, in front of
the planet, this reads:

\begin{eqnarray}
\Delta j_{\rm GI} & = & \int^{t_\infty}_{t_0} \Gamma_{\rm GI}(t)\ dt \\
&=& \int^{\phi_s}_{\phi_\infty} \Gamma_{\rm GI}(\phi)\ (dt/d\phi)\ d\phi,
\end{eqnarray}
where the~GI index stands for ``globally isothermal''.  We then
approximate $(dt/d\phi)= 1/v_\phi \simeq a/2A_px_{s,\rm GI}$, and
since $\Delta j = 2aB_px_{s, \rm GI}$, we have:
\begin{equation}
\label{eq:djFgamma}
4A_pB_px_s^2= \int^{\phi_s}_{\phi_\infty} \Gamma(\phi)\ d\phi
\end{equation}
A similar relation holds in locally isothermal case, with a modified
$x_s = x_{s,\rm GI} + \delta x_s$ and $\Gamma=\Gamma_{\rm GI} +\delta
\Gamma$. The only contributions in $\delta \Gamma$ is the one of the
pressure, assuming that the fluid element follows a similar path in
both cases:
\begin{equation}
\label{eq:27}
\delta \Gamma = \left.\frac{\partial_\phi P}{\Sigma}\right|_{\rm LI} - 
\left. \frac{\partial_\phi P}{\Sigma}\right|_{\rm GI} = 
\delta T \frac{\partial_\phi\Sigma}{\Sigma}, 
\end{equation}
where the LI index stands for ``locally isothermal'', and where
$\delta T = T_{LI} - T_{GI}$.  In equation~(\ref{eq:27}) we have made
use of the fact that $\partial_\phi T$ cancels out.  From equation
(\ref{eq:djFgamma}), we have:
\begin{equation}
\label{eq:28}
8 A_pB_p \delta x_s^F = \int^{\phi_s}_{\phi_\infty} \delta T \frac{\partial_\phi\Sigma}{\Sigma} d\phi.
\end{equation}
We simplify further by assuming ``square'', instantaneous U-turns:
the particles go from $(x_s,\phi_{\infty})$ to $(x_s,\phi_{s})$, then
to $(-x_s,\phi_{s})$, and lastly to $(-x_s,\phi_{\infty})$. This
ensures that $\delta T$ is constant along it, and we are left with:
\begin{eqnarray}
\label{eq:29}
8 A_pB_p \delta x_s^F &\sim & {\delta T (x_s)}\int^{\phi_s}_{\phi_\infty} 
\frac{d\Sigma}{\Sigma} \\
&=& -\tau \frac{x_s}{a} T_0 \frac{\left(\Sigma_s-
    \Sigma_\infty\right)}{\Sigma_c},
\end{eqnarray}
where we have replaced $\delta T$ by its first order expression $
-\tau T_0\ x_s/a$.  Denoting with $\Delta P = P_s-P_\infty$, this
reduces to:
\begin{equation}
\label{eq:30}
\delta x_s^F = -\frac{\tau}{8A_pB_pa}\frac{\Delta P}{\Sigma_c}.
\end{equation}
On the rear of the planet, the sign of $\delta T$ changes, and we have
$\delta x_s^R = -\delta x_s^F$.  The crude estimate presented here
gives an order of magnitude of the intrinsic asymmetry of the
horseshoe region, that gives rise to a torque for the reasons
presented in detail in paper~II.  Since the enthalpy tends to fill the
planet's potential well, we always have $\Delta P > 0$, and the sign
of $\delta x_s^F$ only depends on $\tau$. For example, for a disk with
decreasing temperature ($\tau > 0$), $\delta x_s^F$ is negative, and
the planet feels a negative torque excess from the front region. The
rear region being in turn broader, the planet feels a negative torque
excess from the rear region also, hence a globally negative torque
excess. To quantify this excess, we use the same method as in
paper~II, which yields:

\begin{equation}
\delta \Gamma = 2\Sigma_c\ v_\phi  \delta x_s 
\Delta j\sim  2 \tau \Delta P x_s^2,
\label{eq:torqueFtau}
\end{equation}
where we take $v_{\phi}=2A_px_s$, and $\Delta j = 4B _pa x_s$.

We therefore find a non-isothermal torque excess that scales with the
temperature gradient, and which should have a stronger dependency at
lower softening lengths, since the potential well of the planet is
then deeper, and $\Delta P$ is therefore larger.

\subsection{Comparison to numerical simulations}
\begin{figure}
	\centering
		\plotone{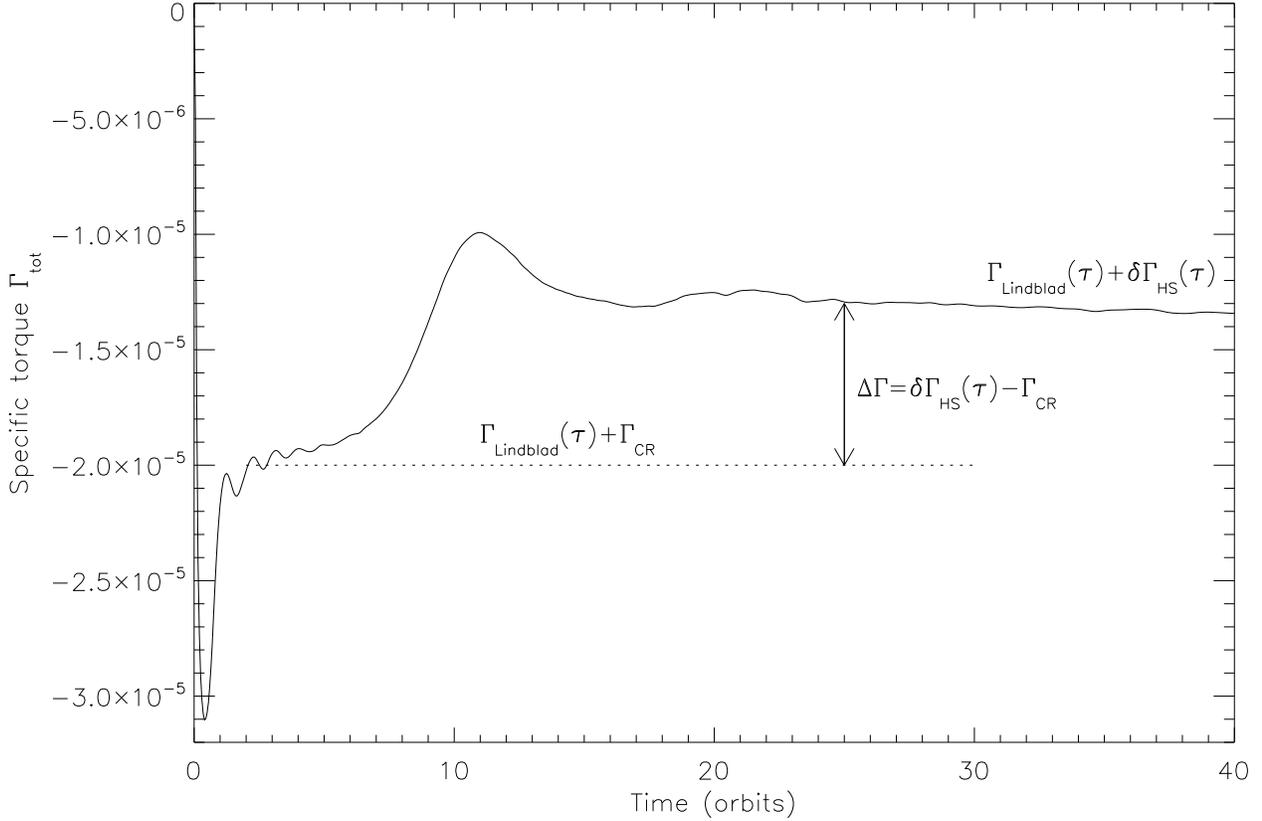}
                \caption{Measurement of $\Delta\Gamma$. For a given
                  run (here, $\tau=-1.35$, $\epsilon=0.3$), we monitor
                  the total torque $\Gamma_{\rm tot}$ as a function of
                  time. The differential Lindblad torque $\Gamma_{\rm
                    Lindblad}$ and the linear co-rotation torque
                  $\Gamma_{\rm CR}$ only take a few orbits to
                  establish, then the horseshoe drag develops, taking
                  place of the linear corotation torque
                  \citep{2009arXiv0901.2265P}. We measure $\Delta
                  \Gamma =\delta \Gamma_{\rm HS}-\Gamma_{\rm CR}$,
                  which, as $\Gamma_{\rm CR}$ does not depends on
                  $\tau$, should trace the dependency of the horseshoe
                  drag on $\tau$. The values of $\Gamma_{\rm
                    Lindblad}+\Gamma_{\rm CR}$ and $\Gamma_{\rm
                    Lindblad}+\delta \Gamma_{\rm HS}$ are measured by
                  averaging $\Gamma_{\rm tot}$ between 2 and 3 orbits
                  on the one hand, and between 30 and 35 on the other
                  hand. Slight changes of theses values did not bring
                  differences.}
	\label{fig:tqftime}
\end{figure}

\begin{figure}
	\centering
		\plotone{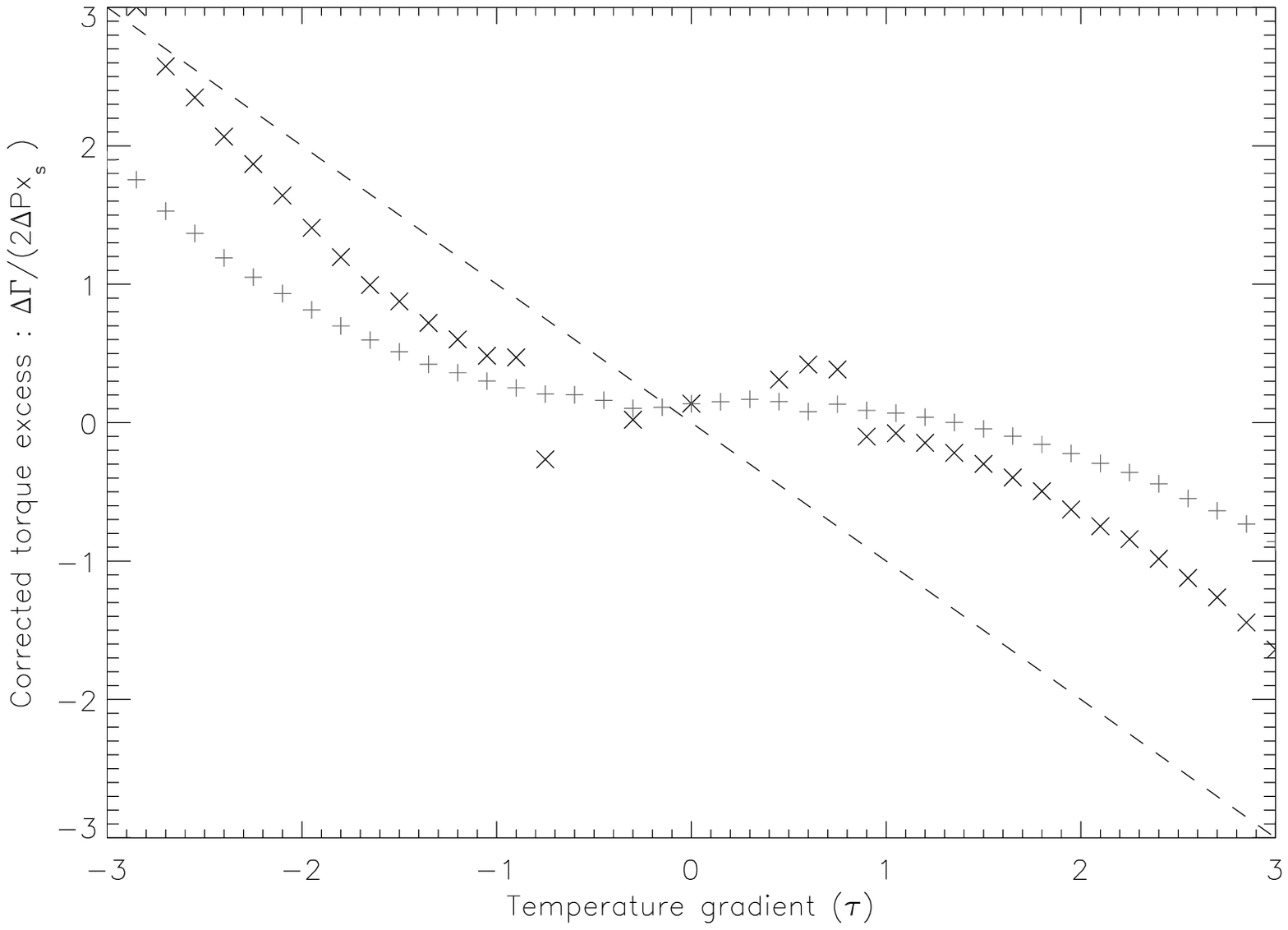}
                \caption{Non-isothermal torque excess. We plot this
                  quantity as a function of $\tau$, for two sets of
                  runs: $\epsilon=0.3$ (X), and $\epsilon=0.5$
                  (+). The dashed line represents our crude estimate
                  of equation~(\ref{eq:torqueFtau}). A proper linear
                  regression fit of the simulated points would yield a
                  slope within $\pm25\%$ of our prediction. The break
                  of continuity that we see for $\epsilon=0.3$ exists
                  also for $\epsilon=0.5$, but is much less apparent,
                  as it takes place at $\tau \approx \pm 0.5$, and
                  results in a much smaller jump of the corrected
                  torque excess.}
	\label{fig:tqftau}
\end{figure}

We performed two sets of $30$ numerical simulations with a varying
temperature slope $\tau$ between $-3$ and $+3$ to investigate the
behavior of the non-isothermal torque excess. The first set uses a
softening length of $\epsilon=0.3$, while the second one uses
$\epsilon=0.5$. Due to a strong generation of vortices when
$\mathcal{V}=0$, we adopted a non-vanishing vortensity gradient, and
we chose $\mathcal{V}=1$. The results of these simulations are
presented in figure \ref{fig:tqftau}. We measured the quantity $\Delta
\Gamma$ as explained in figure \ref{fig:tqftime}. The value of $\Delta
P$ was measured by averaging $\Delta P^F$ and $\Delta P^R$.

There is a poor fit of the numerical data with our crude
estimate. While the slope has the expected value (within an error
range of $\pm25\%$), there are discontinuities in the case
$\epsilon=0.3$ for $|\tau|\sim 1$, which seem to disappear for the
larger value of $\epsilon$. This is linked to the appearance of a
``trapped region'', which we explain in the next section.

\subsection{A note on the softening length}
\label{sec:softening}

\begin{figure}[s]
	\centering 
	\plotone{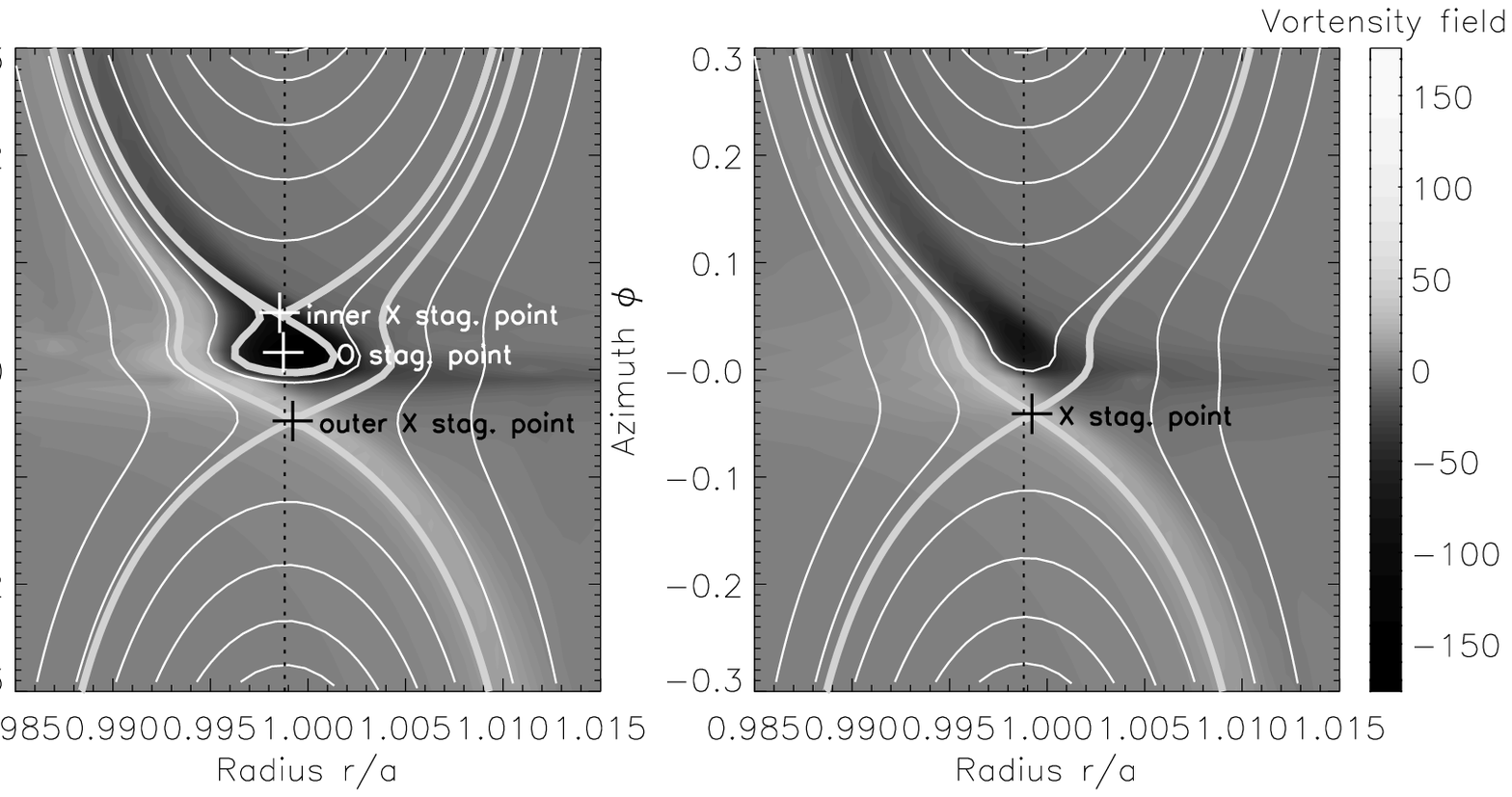} 
        \caption{Appearance of a trapped region. This is a close
          overview of the horseshoe region, at T=13 orbits, close to
          the planet, which is at located at $\phi=0$ and $r=1$. The
          temperature gradient is $\tau=-0.9$. The left panel show the
          $\epsilon=0.3$ case, while the right panel shows the
          $\epsilon=0.5$ case. The thin dotted line represents the
          corotation. The underlying field is the perturbed
          vortensity. The thin white lines are the streamlines, and
          the bold ones represents the separatrices. We clearly see
          the advection of the vortensity created, and the advection
          of negative and positive vortensity to the front region
          (since, here, the outer stagnation point is behind the
          planet). When the inner stagnation point disappears, the
          material executing a U-turn in front of the planet passes
          faster by the planet, hence a smaller creation of negative
          vortensity (hardly seen with the grey scale, but there is a
          difference of about one half), while the creation of
          positive vortensity behind the planet remains of the same
          order.}
	\label{fig:fish}
\end{figure} 

\begin{figure}
	\centering
		\plotone{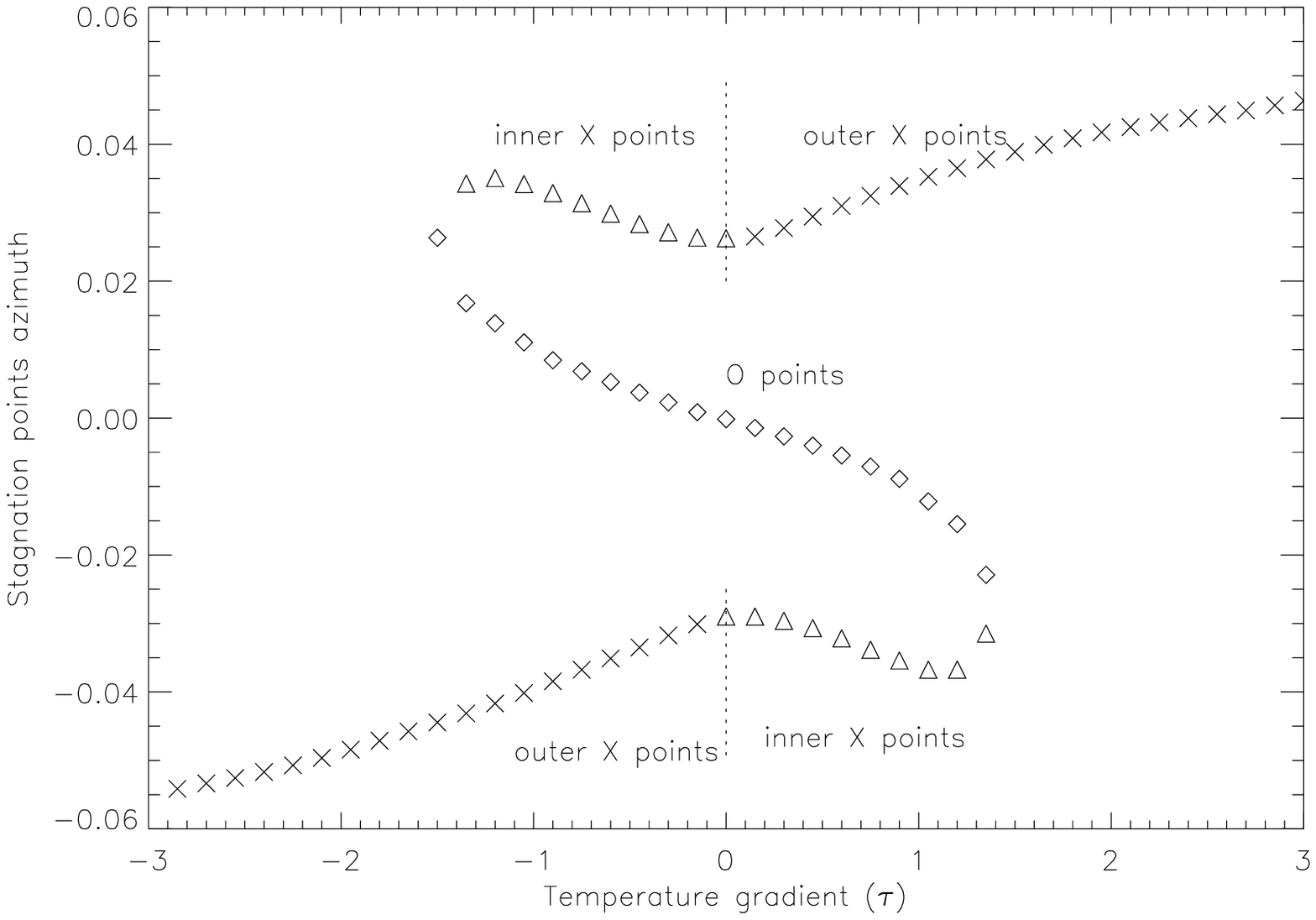}
                \caption{Azimuth of the stagnation point(s) as a
                  function of the temperature gradient, at T=5
                  orbits. The variation of the radial position is much
                  smaller, although it exhibits a reproducible
                  behavior (the front (rear) X stagnation point being
                  always at $r<a$ ($r>a$) ). For $|\tau|<1.5$,
                  there is one O point, and two X-points, as indicated
                  on the figure. At larger times, the inner X-point
                  and the O-point tend to merge, reducing the interval
                  upon which there are three stagnation
                  points. Nevertheless, it remains quite large
                  ($|\tau|<1$) in steady state.}
	\label{fig:stagftau}
\end{figure}

The streamlines in the inner part of the horseshoe region, near the
planet, exhibit different behaviors as a function of the softening
length, as shown in figure \ref{fig:fish}. We can distinguish two
cases:
\begin{itemize}
\item a case where there are three stagnation points, as discussed in
  \cite{mak2006}, with almost the same radius $r_c$. The O-point is
  associated to the planet, being located almost at the same
  azimuth. One of the X-points lies on the separatrices.  We call it
  the outer X-point. Similarly, the other X-point is the one we refer
  to as the inner X-point.
\item a case where there is only one X-point, at the intersection of
  the separatrices.
\end{itemize}
One can continuously go from the first situation to the last one by
varying the softening length.  The inner X-point and the O-point merge
at some point, yielding a one stagnation point configuration.  A lower
softening length favors the existence of an inner X-point, but it is
not the only parameters that determines this property of the flow.
The temperature gradient also plays a role (a smaller gradient favors
the existence of a inner X point, while at large gradient we have a
one stagnation point situation).  We have seen this for two values of
$\epsilon$, and present this for varying values of $\tau$ in figure
\ref{fig:stagftau}. The existence of three stagnation points appears
to be correlated with the discontinuities at $\tau = \pm1$. We comment
on this behavior in the next section.

\subsection{The different regimes in the locally isothermal case}
The production of vortensity takes place near the planet, during the
horseshoe U-turns, where the gradients of pressure (dominated by the
planet's potential) and of temperature are not aligned.  For the sake
of definiteness, we assume that there is a negative temperature
gradient.  Therefore a negative amount of vortensity is produced
behind the planet and a positive amount of vortensity is produced in
front of it. In an idealized situation with a unique X-point centered
on the planet, the negative perturbed vortensity is fully advected to
the rear of the planet, while the positive perturbed vortensity
perturbation is advected towards positive azimuth.  They respectively
yield a positive surface density perturbation behind the planet (for
the reasons exposed in section~\ref{sec:pressure_model}, which are
valid azimuthally far from the planet, where the gradients of
temperature and density are aligned) and a negative perturbation of
surface density in front of the planet (hence a negative
non-isothermal torque excess).

In a realistic situation, however, the fact that the X-point is not
centered on the planet changes this picture. We take the example of
$\tau\sim-1$ (one can refer to figure \ref{fig:fish}b to see a
streamline map that looks similar to this situation). The material at
the rear of the X point is advected to the rear of the planet, as
expected. But some of the material (and the resulting underlying
created positive vortensity) at the front of the X point, but still at
the rear of the planet, is eventually advected to the front of the
planet. This lowers the torque excess exerted on the planet, as a part
of the positive vortensity created at the rear of the planet and
advected to the front cancels out with the negative vortensity created
and advected to the front. This explains why the simulated points in
figure \ref{fig:tqftau} are below our prediction for $\tau <-1$.

As $|\tau|$ decreases, a ``trapped region'' appears, enclosed between
the two X-points, and centered on the O-point.  The material in this
region orbits around the planet, and does not contribute to the
torque. In fact, the created vortensity supported by this material is
trapped here, and even eventually flattens out (under the action of
phase mixing, which requires very long times).  For $\tau<0$, the
azimuth of the O-point is positive, and the material trapped in this
region essentially has a negative created vortensity. Hence, when the
trapped region appears, a significant part of the advected negative
vortensity disappears from the torque, resulting in a smaller torque
excess. This likely explains the discontinuity in
figure~\ref{fig:tqftau} at $\tau \simeq -1$.

Similar arguments hold for the case $\tau>0$ (we then have to change
the sign of the vortensity produced, as well as the sign of the
azimuth of the outer and inner X-points, as can be seen in
figure~\ref{fig:stagftau}).

For $|\tau|>1$, $\Delta \Gamma$ decreases with $\tau$ as announced,
with a slope that is roughly the one predicted in
section~\ref{sec:model_LocIsoT}.  For $|\tau|<1$, however, this slope
is reverted. The reason for this behavior is unclear, and its study
requires an in-depth streamline analysis, as well as a quantitative
estimate of the vortensity produced in the vicinity of the
separatrices, which is beyond the scope of this qualitative
description.

\section{Conclusion}
\label{sec:conclusion}

We have derived an expression for the horseshoe drag that takes into
account the effects of pressure. It yields a result identical to
previous estimates of the horseshoe drag, except that it explicitly
takes into account the material beyond the horseshoe separatrices.
The horseshoe drag therefore accounts for the torque arising from the
whole corotation region, and therefore should deserve to be called
corotation torque.  We find that the horseshoe region is asymmetric,
in a manner that depends on the vortensity gradient. Specifically, the
part in front of the planet is wider for a decreasing vortensity
profile in the disk, while the part behind the planet is more
narrow. This asymmetry can be interpreted as an effect of evanescent
pressure waves, launched from the downstream part of the horseshoe
U-turns, which perturb the upstream region. Using a Bernoulli
invariant, we derived this asymmetry for a globally isothermal disk,
and found it to scale with the vortensity gradient and the disk's
aspect ratio.

The impact of the evanescent waves on the horseshoe width can easily
be interpreted as a consequence of an azimuthal pressure gradient,
whose torque counteracts or adds to the torque required to perform a
U-turn, resulting either in a wider or in a narrower horseshoe region.

This asymmetry does not alter the standard horseshoe drag. Basically,
the supplementary torque excess resulting from the widening of one
side of the horseshoe region is balanced by the torque deficit
resulting of the shrinking of the other side of the horseshoe region.

As the concept of horseshoe drag has proved of increasing importance
for planetary migration in a number of recent works
\citep{trap06,mak2006,pm06,bm08,pp08,2009arXiv0901.2263P,2009arXiv0901.2265P,2008A&A...487L...9K},
it is crucial to adopt a proper estimate of the correct horseshoe
width when evaluating the corresponding horseshoe drag. We find that
the correct mean width of the horseshoe region should be estimated as
given by equation~(\ref{eq:xmean}).  As the asymmetry is relatively
mild for typical vortensity gradients and aspect ratios, a simple
arithmetic mean of the front and rear widths also yields estimates
that are sufficiently accurate.

Our analysis is based on the use of a Bernoulli invariant, which is
made possible by the hypothesis that the disk is globally isothermal,
hence barotropic.  Such an approach is not possible for a locally
isothermal disk.  In such a situation, vortensity is created near the
planet, which gives rise to a supplementary torque. We have presented
a qualitative study of this supplementary torque, whose dependence on
the temperature gradient appears to be intimately linked to the
topology of the flow in the planet vicinity, and which exists in a disk
even without a vortensity gradient. We also provide a crude
estimate of this dependence, which is meant to apply to situations
with only one stagnation point (i.e. at large temperature gradient).
We stress that this supplementary torque is not strong enough to
reverse planetary migration in a disk with a surface density that
decreases outwards as a power law of radius, even with unrealistic
temperature gradients.  Nevertheless, it should be considered whenever
accurate estimates of the migration rate are required, as its
amplitude can amount to about one third of the total horseshoe drag. A
systematic study of these non-isothermal effects, and their
generalization to the three-dimensional case, should be undertaken in
order to provide reliable estimates of this supplementary torque, that
we have called the non-isothermal torque excess.

An important feature of this excess is that it is linked to an edge
effect of the horseshoe drag. The bulk horseshoe drag, which
corresponds to the classical expression, always scales with the
vortensity gradient. Any dependence on another parameter of the disk
(such as the temperature gradient) manifests itself as evanescent
waves excited at the downstream separatrices, where they yield a
vortensity sheet as the most tangible imprint on the flow.  In the
case of a locally isothermal disk, the lack of an invariant and the
strong dependence on the topology of the flow render the situation
very complex and hardly tractable. In the case of an adiabatic flow,
one can exhibit, under certain circumstances, an invariant of the
flow. This allows a rigorous analysis of the horseshoe drag and of the
main properties of the horseshoe region.  This analysis is presented
in paper~II.

\acknowledgments

The numerical simulations performed in this work have been run on the
92 core cluster funded by the program {\em Origine des Plan\`etes et
  de la Vie} of the French {\em Institut National des Sciences de
  l'Univers}. Partial support from the COAST project ({\em
  COmputational ASTrophysics}) of the CEA is also acknowledged. The
authors also wish to thank G. Koenigsberger for hospitality at the
Instituto de Ciencias Fisicas of UNAM, Mexico, and acknowledge partial
support from CONACYT project number 24936.


\begin{thebibliography}{22}

\bibitem[{{Baruteau} \& {Masset}(2008)}]{bm08}
{Baruteau}, C., \& {Masset}, F. 2008, \apj, 672, 1054

\bibitem[{{Goldreich} \& {Tremaine}(1979)}]{gt79}
{Goldreich}, P., \& {Tremaine}, S. 1979, \apj, 233, 857

\bibitem[{{Goldreich} \& {Tremaine}(1980)}]{gt80}
---. 1980, \apj, 241, 425

\bibitem[{{Kley} \& {Crida}(2008)}]{2008A&A...487L...9K}
{Kley}, W., \& {Crida}, A. 2008, \aap, 487, L9

\bibitem[{{Korycansky} \& {Pollack}(1993)}]{1993Icar..102..150K}
{Korycansky}, D.~G., \& {Pollack}, J.~B. 1993, Icarus, 102, 150

\bibitem[{{Masset}(2000{\natexlab{a}})}]{fargo2000}
{Masset}, F. 2000{\natexlab{a}}, \aaps, 141, 165

\bibitem[{{Masset}(2000{\natexlab{b}})}]{fargo2000b}
{Masset}, F.~S. 2000{\natexlab{b}}, in Astronomical Society of the Pacific
  Conference Series, Vol. 219, Disks, Planetesimals, and Planets, ed.
  G.~{Garz{\'o}n}, C.~{Eiroa}, D.~{de Winter}, \& T.~J. {Mahoney}, 75--+

\bibitem[{{Masset}(2001)}]{masset01}
{Masset}, F.~S. 2001, \apj, 558, 453

\bibitem[{{Masset}(2002)}]{masset02}
---. 2002, \aap, 387, 605

\bibitem[{{Masset} {et~al.}(2006{\natexlab{a}}){Masset}, {D'Angelo}, \&
  {Kley}}]{mak2006}
{Masset}, F.~S., {D'Angelo}, G., \& {Kley}, W. 2006{\natexlab{a}}, \apj, 652,
  730

\bibitem[{{Masset} {et~al.}(2006{\natexlab{b}}){Masset}, {Morbidelli}, {Crida},
  \& {Ferreira}}]{trap06}
{Masset}, F.~S., {Morbidelli}, A., {Crida}, A., \& {Ferreira}, J.
  2006{\natexlab{b}}, \apj, 642, 478

\bibitem[{{Masset} \& {Papaloizou}(2003)}]{mp03}
{Masset}, F.~S., \& {Papaloizou}, J.~C.~B. 2003, \apj, 588, 494

\bibitem[{{Paardekooper} \&
  {Papaloizou}(2009{\natexlab{a}})}]{2009arXiv0901.2263P}
{Paardekooper}, S.~., \& {Papaloizou}, J.~C.~B. 2009{\natexlab{a}}, \mnras, 394, 2297

\bibitem[{{Paardekooper} \& {Mellema}(2006)}]{pm06}
{Paardekooper}, S.-J., \& {Mellema}, G. 2006, \aap, 459, L17

\bibitem[{{Paardekooper} \& {Papaloizou}(2008)}]{pp08}
{Paardekooper}, S.-J., \& {Papaloizou}, J.~C.~B. 2008, \aap, 485, 877

\bibitem[{{Paardekooper} \&
  {Papaloizou}(2009{\natexlab{b}})}]{2009arXiv0901.2265P}
{Paardekooper}, S.~J., \& {Papaloizou}, J.~C.~B. 2009{\natexlab{b}}, \mnras, 394, 2283

\bibitem[{{Tanaka} {et~al.}(2002){Tanaka}, {Takeuchi}, \& {Ward}}]{tanaka2002}
{Tanaka}, H., {Takeuchi}, T., \& {Ward}, W.~R. 2002, \apj, 565, 1257

\bibitem[{{Ward}(1986)}]{ww86}
{Ward}, W.~R. 1986, Icarus, 67, 164

\bibitem[{{Ward}(1988)}]{WW88}
---. 1988, Icarus, 73, 330

\bibitem[{{Ward}(1989)}]{1989ApJ...336..526W}
---. 1989, \apj, 336, 526

\bibitem[{{Ward}(1991)}]{wlpi91}
{Ward}, W.~R. 1991, in Lunar and Planetary Institute Conference Abstracts,
  1463--+

\bibitem[{{Ward}(1992)}]{wlpi92}
{Ward}, W.~R. 1992, in Lunar and Planetary Institute Conference Abstracts,
  1491--+

\end{thebibliography}
\end{document}